\documentclass[journal]{IEEEtran}

\usepackage{color,soul,framed} 

\usepackage[pdftex]{graphicx}
\graphicspath{{../figures/}{../jpeg/}}
\DeclareGraphicsExtensions{.pdf,.jpeg,.png}

\usepackage[cmex10]{amsmath}
\usepackage{amssymb}
\usepackage{nccmath}
\usepackage{array}
\usepackage{mdwmath}
\usepackage{mdwtab}
\usepackage{eqparbox}
\usepackage{url}
\usepackage{hyperref}
\usepackage{multirow}
\usepackage{makecell}
\usepackage{bbding}
\usepackage{subfigure}
\newcolumntype{P}[1]{>{\centering\arraybackslash}p{#1}}

\begin{document}
\bstctlcite{IEEEexample:BSTcontrol}
    \title{Multi-Metric Optimization using Generative Adversarial Networks for Near-End Speech Intelligibility Enhancement}
  \author{Haoyu~Li,~\IEEEmembership{Student Member,~IEEE,} and
      Junichi~Yamagishi,~\IEEEmembership{Senior Member,~IEEE}

  \thanks{Manuscript received April 9, 2021; revised August 13, 2021; accepted September 5, 2021. This work was supported in part by JST CREST VoicePersonae project under Grant JPMJCR18A6, Japan, in part by MEXT KAKENHI Grants 16H06302, 17H04687, 18H04120, 18H04112, 18KT0051, and 19K24372, Japan, and in part by SOKENDAI (The Graduate University for Advanced Studies), Japan. \textit{(Corresponding author: Haoyu Li.)}}

  \thanks{Haoyu Li and Junichi Yamagishi are with the National Institute of Informatics, and with the Department of Informatics, SOKENDAI, Tokyo 101-8340, Japan (e-mail: \{haoyuli, jyamagis\}@nii.ac.jp).} 
  }

\markboth{IEEE/ACM TRANSACTIONS ON AUDIO, SPEECH, AND LANGUAGE PROCESSING, VOL. XX, NO. XX, XX 2021
}{Li \MakeLowercase{\textit{et al.}}: Multi-Metric Optimization for Near-End Speech Intelligibility Enhancement}

\maketitle

\begin{abstract}
The intelligibility of speech severely degrades in the presence of environmental noise and reverberation. 
In this paper, we propose a novel deep learning based system for modifying the speech signal to increase its intelligibility under the equal-power constraint, i.e., signal power before and after modification must be the same.
To achieve this, we use generative adversarial networks (GANs) to obtain time-frequency dependent amplification factors, which are then applied to the input raw speech to reallocate the speech energy. 
Instead of optimizing only a single, simple metric, we train a deep neural network (DNN) model to simultaneously optimize multiple advanced speech metrics, including both intelligibility- and quality-related ones, which results in notable improvements in performance and robustness. 
Our system can not only work in non-real-time mode for offline audio playback but also support practical real-time speech applications. 
Experimental results using both objective measurements and subjective listening tests indicate that the proposed system significantly outperforms state-of-the-art baseline systems under various noisy and reverberant listening conditions. 

\end{abstract}

\begin{IEEEkeywords}
speech intelligibility, generative adversarial networks, multi-metric optimization
\end{IEEEkeywords}

\IEEEpeerreviewmaketitle

\section{Introduction}

\IEEEPARstart{R}{eal-life} speech communication, such as mobile telephony and public-address announcement, usually occurs in noisy and reverberant environments. 
These challenging environments severely degrade speech intelligibility, resulting in stressful listening or even non-understanding for listeners.
Since noise sources are physically present in the near-end listener side, typical speech enhancement methods (e.g., \cite{boll1979suppression, ephraim1985speech, xu2014regression}) which recover the clean speech from the noisy input, however, cannot be applied in such scenarios.
As an alternative, there are many other methods aiming to modify the speech signal only to improve its intelligibility when exposed to noise and reverberation.
In this paper, we refer to this task as near-end speech intelligibility enhancement. 

Numerous algorithms for near-end intelligibility enhancement have been studied over the past decade (e.g., \cite{nathwani2016formant, optSII, tang2012optimised, ZorilaKS12, hendriks2015optimal, tang2018learning, Chermaz2020}). 
In particular, the 1st and 2nd Hurricane challenges \cite{cooke2013intelligibility, Rennies2020} summarized many effective algorithms and conducted comprehensive comparisons for each, providing remarkable reference value for researchers.

To increase intelligibility, many modification algorithms were designed on the basis of expert knowledge. 
For example, one algorithm called \mbox{SSDRC} \cite{ZorilaKS12} empirically sharpens the formant information and reduces the envelope variations of a speech signal, which leads to significant intelligibility improvement. 
Another example is a method called ASE \cite{Chermaz2020}, which maximizes intelligibility through certain audio manipulations, such as frequency-band decomposition and dynamic range compression, on the basis of sound engineering knowledge. 
Although these algorithms clearly improve speech intelligibility, they are dependent on domain experts' subjective experiences, thus still leaving room for improvement.
These algorithms also consist only of non-parametric speech modifications; therefore, they cannot adapt well to the changing environments.

Inspired by human speech production characteristics, some algorithms (e.g., \cite{nathwani2017speech, lopez2017speaking, seshadri2019augmented}) aim to convert normal speech to Lombard speech \cite{lane1971lombard}, which is naturally produced by speakers with increased vocal effort for higher intelligibility. 
To achieve speaking style conversion, most algorithms rely on vocoder-based analysis-and-synthesis techniques, where vocoder features are transformed to fit in the Lombard style. For example, \mbox{Seshadri} \MakeLowercase{\textit{et al.}} \cite{seshadri2019augmented} modified Mel-generalized cepstrum coefficients \cite{mgcc1994} of input speech to generate the Lombard-style speech by using log-domain pulse model vocoder \cite{degottex2017log}.
However, using such a parametric vocoder inevitably degrades the converted speech quality. 
It was also found that even natural Lombard speech could only produce very limited intelligibility gains under low signal-to-noise ratio (SNR) conditions \cite{cooke2013intelligibility}. 
Consequently, the performance of such Lombard-inspired algorithms is still far from satisfactory.

Another group of algorithms were developed through optimizing certain objective intelligibility metrics. 
The basic concept is to modify the input speech in such a way as to maximize a target intelligibility metric under a known noise condition. 
For example, some algorithms (e.g., \cite{optSII} and \cite{hendriks2015optimal}) were proposed to maximize the speech intelligibility index (SII) \cite{american1997american}.
Another group \cite{tang2012optimised, tang2018learning, valentini2014intelligibility} optimizes a glimpse-based intelligibility metric \cite{cooke2006glimpsing}. 
These algorithms show promising results and do not rely on expert knowledge. Nevertheless, their performance still falls behind state-of-the-art algorithms such as SSDRC in subjective tests, as previously reported in \cite{cooke2013intelligibility}. 
This is because the objective metrics (e.g., SII) optimized within the above algorithms are relatively simple and inaccurate, i.e., they are not highly correlated with subjective intelligibility across different types of noise and other signal degradations \cite{van2018evaluation}.
Also, optimizing only a single target usually causes sub-optimality in another metrics, therefore limiting performance. 
Several advanced intelligibility metrics have recently been proposed and showed good results \cite{KATES2020, van2017instrumental}.
However, it is still difficult to find closed-form solutions for optimizing these metrics due to mathematical complexities.
Although numerical methods, such as gradient descent\footnote{Gradient descent algorithm is inapplicable to the non-differentiable target metrics, where the gradients cannot be calculated.} and the genetic algorithm \cite{holland1992adaptation}, can simultaneously optimize multiple complex metrics, their optimization schemes are based on offline iterative updates, and thus not suitable for real-time online applications.

Inspired by progresses in black-box function optimization \cite{fu2019metricgan, kawanaka2020stable}, we previously proposed a generative adversarial network (GAN)-based system \cite{Li2020} for near-end intelligibility enhancement. 
The system was composed of a generator that enhances the intelligibility of input speech and a discriminator that acts as a learned surrogate of evaluation metrics to guide the training scheme of the generator.
We found that multiple intelligibility metrics could be effectively optimized with such a system. Nevertheless, this was still a non-causal system, and its generalization capability for unseen environments had not yet been extensively investigated. 

In this paper, we propose a causal and light-weight system as an extension to our earlier system \cite{Li2020}.
We substitute the original bidirectional long-short term memory (BLSTM) with causal convolution. To decrease the number of model parameters, we operate the speech signal on compact frequency-domain bands with equivalent rectangular bandwidth (ERB) scale resolution \cite{glasberg1990derivation}.
We use a GAN-based optimization scheme to jointly maximize not only multiple intelligibility metrics but also quality metrics for improved speech quality. 
We comprehensively evaluate the system's performance under different conditions with unseen noises and reverberations. 
Our experiments show that the improved system significantly increases the intelligibility and quality of speech over our original system \cite{Li2020} with far less parameters. 
Moreover, it also outperforms the state-of-the-art SSDRC baseline in both objective and subjective evaluations.

The rest of the paper is structured as follows. Section~\ref{sec:sec2} describes the application scenario and formulates the problem of near-end speech intelligibility enhancement. Section~\ref{sec:sec3} gives details on the proposed system. Section~\ref{sec:sec4} presents the experimental setup and results. We conclude this paper in Section~\ref{sec:sec5}.

\section{Scenario Description and Problem Formulation}
\label{sec:sec2}

\begin{figure}
    \begin{center}
        \includegraphics[width=0.48\textwidth]{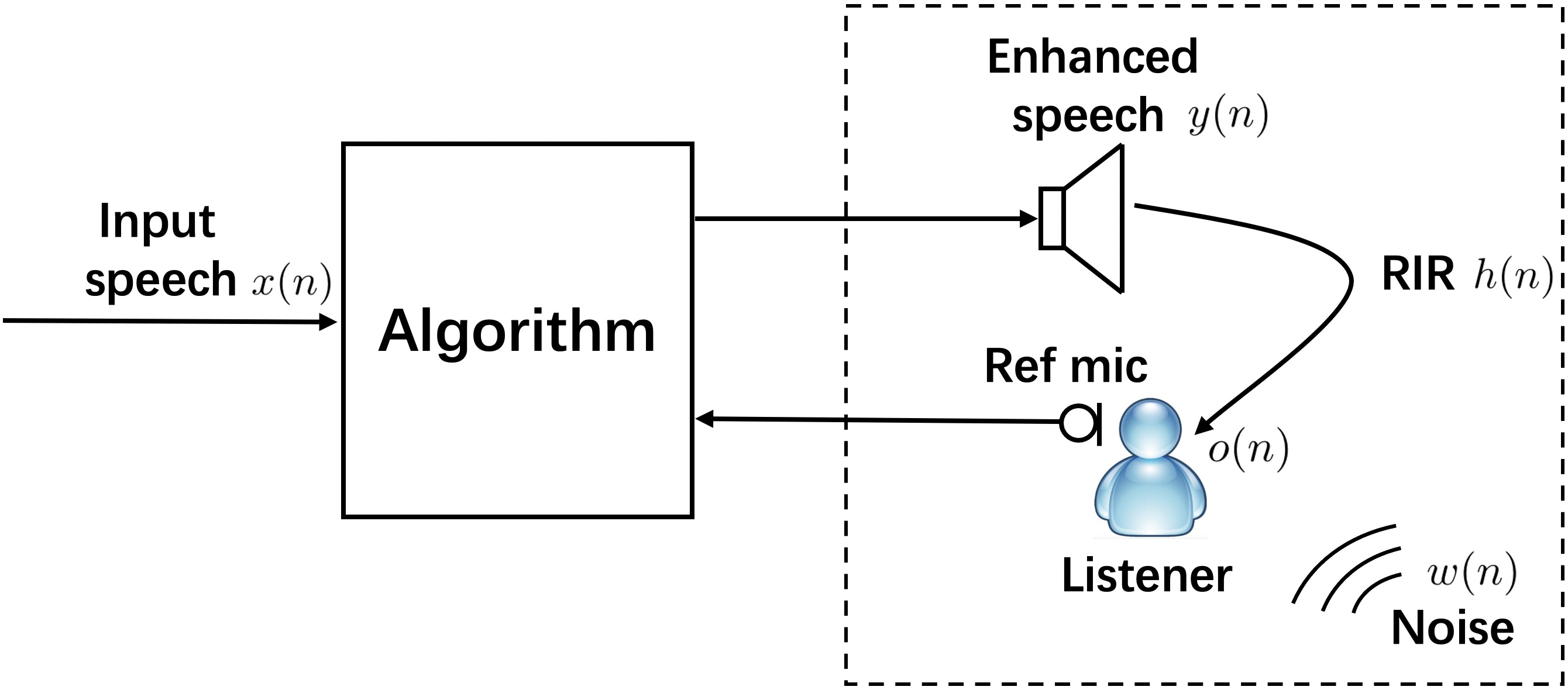}\\
        \caption{Real-life scenario of near-end speech intelligibility enhancement.}
        \label{fig:scenario}
    \end{center}
    \vspace{-1mm}
\end{figure}

Figure~\ref{fig:scenario} depicts an application scenario of near-end speech intelligibility enhancement. Let $x(n)$ be the input speech signal with sampling index $n$. An algorithm is applied to modify $x(n)$, and then the enhanced signal $y(n)$ is output and played via a loudspeaker in a noisy and reverberant environment. The signal $o(n)$ observed by the near-end listener can thus be represented as
\begin{equation}
    \label{eq: observed_signal}
    o(n) = y(n) * h(n) + w(n),
\end{equation}
where $*$ denotes the convolution operation, $h(n)$ is the room impulse response (RIR)\footnote{Loudspeaker response is integrated into the RIR for simplicity.}, and $w(n)$ is the additive noise disturbance. 
We further consider a common scenario in which the noise properties of $w(n)$ can be measured using a noise tracking algorithm via a reference microphone, such as the phone microphone for mobile telephony.
On the other hand, we disregard the effect of reverberation, i.e., $h(n)$, since in practice it is difficult to estimate reverberation parameters in the presence of additive noise.
With these assumptions, our target is thus to develop a system that transforms $x(n)$ into $y(n)$ to improve the intelligibility of $o(n)$ under a known noise condition.

More specifically, speech modification is carried out to redistribute the speech energy over time and frequency. Let $X(m,k)$ be the short-time Fourier transform (STFT) spectrogram of the raw signal $x(n)$, with the frame index $m$ and frequency index $k$. We divide and group the frequency bins into the ERB-scaled bands \cite{glasberg1990derivation} using triangular filter banks with the peak response being at the boundary between bands. Therefore, the input speech energy within one ERB band (indexed by band $i$ at frame $m$) is given by
\begin{equation}
\label{eq:speech_energy}
    E_x(m, i) = \sum\limits_{k}g_i(k)|X(m,k)|^2, 
\end{equation}
where $i\in\{1, 2, \cdots, I\}$ with $I$ the total number of ERB-scaled bands, and $g_i(k)$ is the amplitude of the $i$-th triangular band at the \mbox{$k$-th} frequency bin. 
Similarly, we denote the spectrogram and energy band of noise as $W(m,k)$ and $E_w(m, i)$, respectively. 
The modified speech energy within one ERB band can be represented as $\alpha^2(m,i)E_x(m,i)$, where $\alpha(m,i)$ are the amplification factors that redistribute the speech energy across time and frequency bands. Due to the equal-power constraint, we also have the following equation with respect to $\alpha(m,i)$:
\begin{equation}
    \label{eq:power_constraint}
    \sum\limits_{m,i}\alpha^2(m,i)E_x(m,i)=\sum\limits_{m,i}E_x(m,i).
\end{equation}
Next, the interpolated amplification factors applied to each frame $m$ and frequency bin $k$ are obtained by
\begin{equation}
    \label{eq:interpolation}
    \hat{\alpha}^2(m,k)= \sum\limits_{i}g_i(k)\alpha^2(m,i).
\end{equation}
They are then multiplied with the input spectrogram $X(m,k)$ to produce the enhanced spectrogram $\hat{\alpha}(m,k)X(m,k)$, which is subsequently converted to the enhanced signal $y(n)$ through the inverse STFT.

Instead of relying on expert knowledge to design an algorithm, we select several objective intelligibility and quality metrics as our optimization targets. We will further introduce the selected metrics in Section~\ref{sec3:sub1}. On the basis of the above discussion, we now reformulate the problem as follows. Given the noise estimation (in the form of $E_w(m, i)$) and the constraint of Equation~(\ref{eq:power_constraint}), \textit{our target is to find the amplification factors $\alpha(m,i)$ per time frame and ERB band to optimize the objective metrics of interest.}

\section{Proposed System}
\label{sec:sec3}
In this section, we introduce our proposed GAN-based system to jointly optimize multiple speech metrics for improved intelligibility. 

\subsection{Target Speech Metrics}
\label{sec3:sub1}
Objective metrics are used to measure the intelligibility of speech distorted by noise and reverberation. Very recently, \mbox{Van Kuyk} \MakeLowercase{\textit{et al.}} \cite{van2018evaluation} extensively tested the accuracy of many of these metrics by comparing their correlation coefficients with listening test scores. We accordingly selected the top three intelligibility metrics to build up and evaluate our proposed system. Their notations and brief descriptions are given as follows.
\begin{itemize}
    \item \textbf{SIIB:} Speech intelligibility in bits (SIIB) \cite{van2017instrumental} computes an estimation of the information shared between the clean and distorted speech signals in bits per second. Since an SIIB score relates to the signal duration, all stimuli are either repeated or truncated to have a consistent duration of 20 seconds when using SIIB, producing scores in the range of $[0, +\infty)$.
    \item \textbf{HASPI:} Hearing-aid speech perception index (HASPI) \cite{kates2014hearing} estimates the intelligibility loss through the analysis of cepstral correlation and auditory coherence within an auditory model.
    To obtain the intelligibility score, we use a modified variant proposed in its recent improved version \cite{KATES2020}, where the final score, within the range of $[0, +\infty)$, is calculated as a weighted sum of the modulation filter outputs.
    \item \textbf{ESTOI:} Extended short-time objective intelligibility (\mbox{ESTOI}) \cite{jensen2016algorithm} measures intelligibility by computing the correlation between the spectral of clean and distorted speech. ESTOI score ranges from 0 to 1.
\end{itemize}

In addition to these intelligibility metrics, we also selected the following two state-of-the-art quality metrics.
\begin{itemize}
    \item \textbf{PESQ:} Perceptual evaluation of speech quality (PESQ) is a metric defined in ITU-T recommendation P.862 \cite{rix2001perceptual} for automated assessment of speech quality. PESQ score ranges from -0.5 to 4.5.
    \item \textbf{ViSQOL:} Virtual speech quality objective listener (ViSQOL) is an objective metric released by Google \cite{sloan2017objective, chinen2020visqol} for perceived audio quality. ViSQOL score ranges from 1 to 5.
\end{itemize}
These two quality metrics are incorporated as the optimization targets to compensate for the quality loss caused by intelligibility-enhancing modifications. 

All the above-mentioned metrics are so-called intrusive (or full-reference) models, which require a clean speech signal as the reference to predict an intelligibility or quality score for distorted speech. 
Although they can achieve high correlations with human subjective evaluations, they are too complex and mathematically intractable to handle. Particularly for a deep neural network (DNN) model, we cannot directly use such metrics as the training criteria since most of them are non-differentiable\footnote{ESTOI and PESQ metrics are technically differentiable under certain approximations, which have been studied in \cite{fu2018end} and \cite{kim2019end}, respectively.}. To overcome this obstruction, we thus introduce the following GAN model into our system.

\subsection{System Overview}
\label{sec3:sub2}
\begin{figure}[t]
    \centering
    \subfigure[Training process of discriminators]{
        \includegraphics[height=65mm,width=45mm]{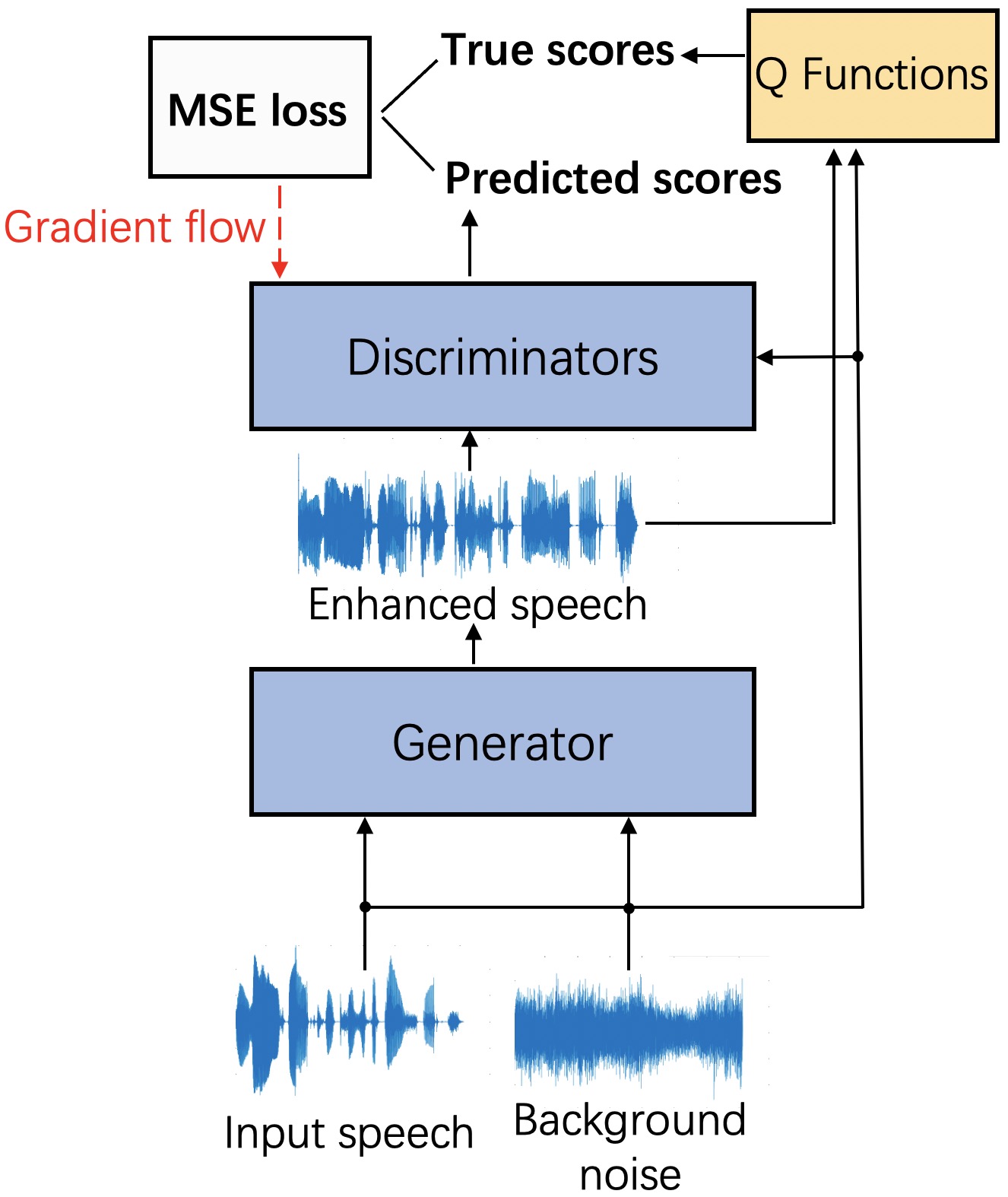}
        }
    \subfigure[Training process of generator]{
        \includegraphics[height=64mm,width=37.5mm]{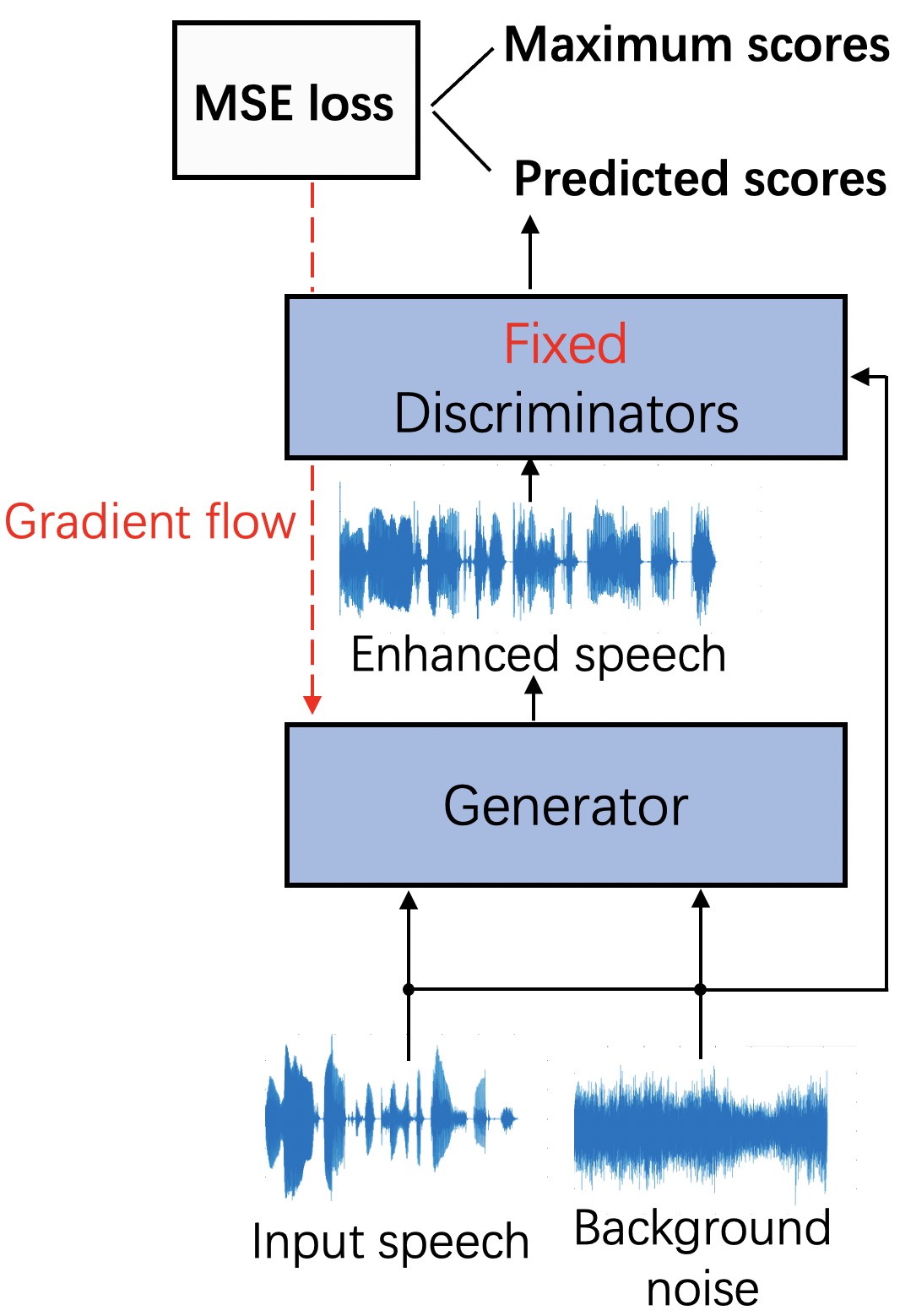}
        }     
    \caption{Diagram of the GAN model of the proposed system for near-end speech intelligibility enhancement.}
    \label{fig:system_diagram}
    \vspace{-1mm}
\end{figure}

Figure~\ref{fig:system_diagram} shows the diagram of the GAN model of our proposed system. It is composed of a generator ($G$), an intelligibility discriminator ($D_{\rm{int}}$), and a quality discriminator ($D_{\rm{qua}}$).
The $G$ receives the input speech $x$ and noise $w$ and then outputs the enhanced speech $y=G(x,w)$, where we omit sampling index $n$ from this point forward.
Next, $D_{\rm{int}}$ and $D_{\rm{qua}}$ predict the intelligibility and quality scores of the enhanced speech, respectively. The predicted scores of the discriminators are expected to be close to the true scores calculated from the target objective metrics. Compared with those original metrics, which are quite complex, the gradients of DNN-based discriminators can be easily computed and back-propagated to $G$. Therefore, with the guidance of $D_{\rm{int}}$ and $D_{\rm{qua}}$, $G$ can be effectively trained to optimize the learned metrics of interest.

\begin{figure*}[t]
    \centering
    \subfigure[$G$ architecture]{
        \includegraphics[height=30mm,width=93mm]{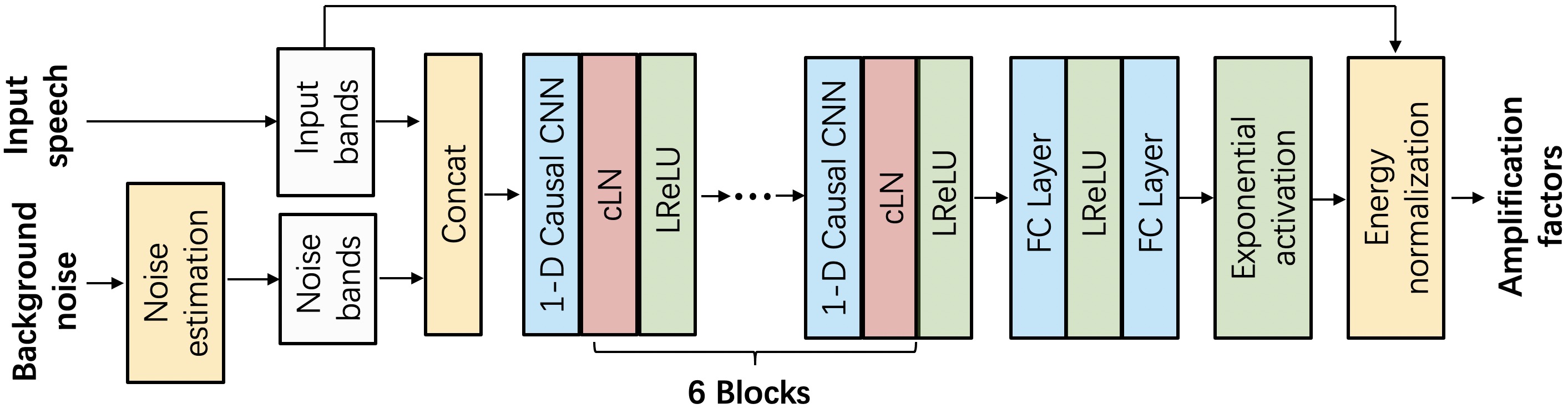}
        }
    \subfigure[$D_{\rm{qua}}$ architecture]{
        \includegraphics[height=28mm,width=82mm]{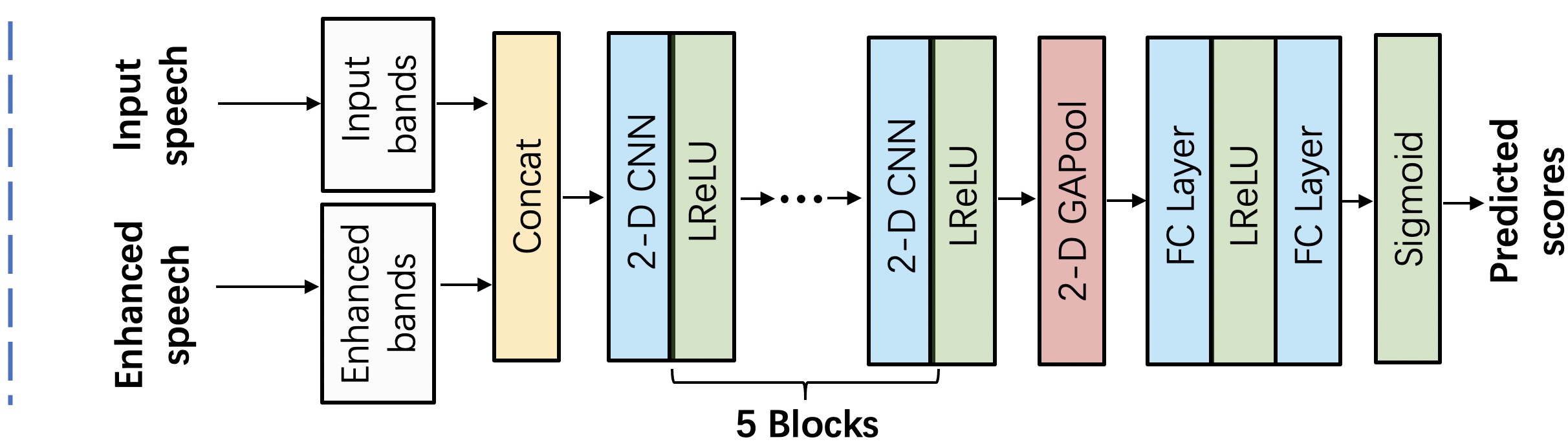}
        }     
    \caption{Network architectures of our GAN model. Concat denotes concatenation operation. We set slope $=0.3$ for all LReLU activations used in experiments.}
    \label{fig:network_architecture}
    \vspace{-2mm}
\end{figure*}

More specifically, we now explain the training process of $D_{\rm{int}}$ in detail. As shown in Fig.~\ref{fig:system_diagram}(a), to predict the intelligibility scores, $D_{\rm{int}}$ takes three inputs: (1) the enhanced speech $G(x,w)$; (2) undistorted clean speech $x$; and (3) background noise $w$. We introduce the so-called $Q$ functions to represent the target metrics to be modelled (described in Section~\ref{sec3:sub1}), with $Q_{\rm{int}}(.)$ the functions for intelligibility metrics and $Q_{\rm{qua}}(.)$ for quality metrics.
Moreover, the signal example $\hat{y}$, which is pre-enhanced using other reference algorithms (e.g., SSDRC \cite{ZorilaKS12} and OptSII \cite{optSII}), is also fed into $D_{\rm{int}}$ in the training.
As demonstrated in our earlier study \cite{Li2020}, learning such additional examples can stabilize the training process and improve performance.
Given all the above notations, the loss function of $D_{\rm{int}}$ is represented as follows:
\begin{equation}
    \begin{aligned}
        \mathcal{L}_{D}^{int} = \mathbb{E}_{x,w} \{ & [D_{\rm{int}} (G(x,w),x,w) - Q_{\rm{int}}(G(x,w),x,w)]^2\\
        & + [D_{\rm{int}}(\hat{y},x,w) - Q_{\rm{int}}(\hat{y},x,w)]^2 \}.
    \end{aligned}    
    \label{eq:L_D_intel}
\end{equation}
By minimizing $\mathcal{L}_{D}^{int}$, $D_{\rm{int}}$ is encouraged to accurately predict the intelligibility scores. 
Similarly, we can represent the loss function of $D_{\rm{qua}}$ as Equation~(\ref{eq:L_D_qua}).
\begin{equation}
    \begin{aligned}
    \mathcal{L}_{D}^{qua} = \mathbb{E}_{x,w} \{ & [D_{\rm{qua}} (G(x,w),x) - Q_{\rm{qua}}(G(x,w),x)]^2\\
        & + [D_{\rm{qua}}(\hat{y},x) - Q_{\rm{qua}}(\hat{y},x)]^2 \}
    \end{aligned}    
    \label{eq:L_D_qua}
\end{equation}
Note that different from $D_{\rm{int}}$, $D_{\rm{qua}}$ takes only two inputs: the enhanced speech $G(x,w)$ and clean reference speech $x$. This is because we have $D_{\rm{qua}}$ focus on measuring the quality of enhanced speech rather than the noisy observed speech.

Figure~\ref{fig:system_diagram}(b) illustrates the $G$ training process. We first fix the parameters of $D_{\rm{int}}$ and $D_{\rm{qua}}$, and then apply the back-propagated gradients to update $G$ to maximize the predicted intelligibility and quality scores. In order to increase the predicted scores as much as possible, we use the following loss function:
\begin{equation}
    \begin{aligned}
    \mathcal{L}_{G} = \mathbb{E}_{x,w} \{ & [D_{\rm{int}} (G(x,w),x,w) - t_{\rm{int}}]^2\\
        & + \lambda [D_{\rm{qua}} (G(x,w),x) - t_{\rm{qua}}]^2 \},
    \end{aligned}    
    \label{eq:G_update}
\end{equation}
where $t_{\rm{int}}$ and $t_{\rm{qua}}$ denote the maximum scores of the selected intelligibility and quality metrics, respectively, and $\lambda$ is a hyper-parameter controlling the weight of speech quality to compensate for the quality degradation caused by intelligibility-enhancing modifications.

The generator ($G$) and discriminators ($D_{\rm{int}}$ and $D_{\rm{qua}}$) are trained alternatively. At one training step, $D_{\rm{int}}$ and $D_{\rm{qua}}$ are trained individually with their corresponding loss functions, i.e., $\mathcal{L}_{D}^{int}$ and $\mathcal{L}_{D}^{qua}$. At the next training step, we fix the discriminators and only train $G$ by minimizing loss $\mathcal{L}_{G}$. By this means, $G$ can be effectively trained to optimize multiple advanced speech metrics, and the intelligibility of the enhanced speech (output by $G$) can be greatly improved and without too much quality degradation.

\subsection{Network Architectures}
\label{sec3:sub3}
The details of the network architectures are given in Fig.~\ref{fig:network_architecture}.
\subsubsection{Generator}
\label{sec3:sub3:sub1}
The input features for $G$ are extracted from input speech and background noise. Specifically, the speech signal is transformed into the features containing 64 ERB-derived bands per time frame using Equation~(\ref{eq:speech_energy}). 
There are two advantages for choosing features in the form of ERB-scaled bands rather than the raw frequency bins: (1) ERB-filterbank groups several perceptually-similar frequency bins into one band, producing a more robust feature\footnote{Filterbank-based grouping operations are also implemented as front-end processing in many intelligibility metrics such as SIIB\cite{van2017instrumental} and ESTOI \cite{jensen2016algorithm}.}; and (2) The number of ERB bands is less than that of frequency bins, which can reduce the dimensions of the input and output features, resulting in a smaller model size.
For background noise, we use the improved minima controlled recursive averaging algorithm (IMCRA) \cite{cohen2003noise} to estimate noise power spectral density (PSD) $W^2(m,k)$, and then similarly extract 64 ERB bands as the noise features. These two features are then concatenated, resulting in a 128-channel feature vector and passed on to the following networks. 

For network design, we choose the 1-D convolutional neural network (CNN) as the backbone for $G$ due to the following reasons: (1) temporal convolution (1-D CNN with filter across time axis) has shown powerful modeling ability and been widely used in speech enhancement \cite{luo2019conv, pandey2019tcnn} and synthesis \cite{ren2020fastspeech}; and (2) the 1-D CNN is suited for real-time applications due to its low computational complexity.

As shown in Fig.~\ref{fig:network_architecture}(a), $G$ consists of six blocks of causal 1-D CNN each with cumulative layer normalization (cLN) \cite{luo2019conv} and LeakyReLU activation (LReLU) \cite{xu2015empirical}. The kernel size and output channels are set to (5, 256), (7, 256), (7, 256), (7, 256), (7, 256), and (5, 64), respectively. Two 64-node fully connected (FC) layers are subsequently followed by the last CNN block. The element-wise exponential activation function is then applied as follows:
\begin{equation}
    output = \exp{(3*\tanh{(u)})},
    \label{eq:exp_act}
\end{equation}
where $u$ is the result of the last FC layer, and the scale range of Equation~(\ref{eq:exp_act}) is approximately 0.05 to 20. 
The 64-dimensional output vector serves as the raw (non-normalized) amplification factors $\alpha(m,i)$, which redistribute the speech energy across time and frequency bands: the speech energy $E_x(m,i)$ (at frame $m$ within band $i$) is boosted when $\alpha(m,i)>1$; otherwise, suppressed. Furthermore, we add an energy normalization layer where the raw amplification factors are multiplied by a global scale factor $\gamma$ in order to satisfy the equal-power constraint of Equation~(\ref{eq:power_constraint}). Finally, the normalized $\alpha(m,i)$ are applied to reconstruct the enhanced speech signal, as described in Section~\ref{sec:sec2}.

Except the last energy normalization operation, all layers in $G$ are designed with causal configurations, which can run without dependencies of the future values of the signal. Moreover, $G$ is a light-weight model containing only around 2.1M parameters. It performs intelligibility enhancement very fast at the frame level, allowing for practical real-time speech applications. We will further discuss the extensions to real-time execution of our proposed system in Section~\ref{sec:sec4:realtime}. 

\subsubsection{Discriminators}
\label{sec3:sub3:sub2}
Figure~\ref{fig:network_architecture}(b) gives the detailed architecture of $D_{\rm{qua}}$. It takes two types of ERB bands as input features: the unmodified input speech bands and enhanced bands. The $D_{\rm{qua}}$ is composed of five layers of 2-D CNN with the following kernel size and number of channels: [(1, 1), 8], [(3, 3), 16], [(5, 5), 32], [(7, 7), 48], and [(9, 9), 64], each with LReLU activation. A 2-D global average pooling (GAPool) \cite{lin2013network} is added to the last CNN block to produce a fixed 64-dimensional output vector, which is then followed by an FC layer with 64 LReLU nodes. The last FC layer with sigmoid activation predicts the scores of modelled quality metrics, i.e., PESQ and ViSQOL. Thus, the number of nodes are accordingly set to 2. Similar to our previous study \cite{Li2020}, we apply spectral normalization with 1-Lipschitz continuity \cite{miyato2018spectral} to all the layers used in $D_{\rm{qua}}$ to stabilize the training process.

For $D_{\rm{int}}$, it shares the same network architecture with $D_{\rm{qua}}$, except the inputs are changed to 3-channel features, i.e., (input, enhanced, noise), which requires an additional input of the estimated noise bands. Besides, the output nodes of $D_{\rm{int}}$ are set to 3, corresponding to the three intelligibility metrics to be modelled: SIIB, HASPI, and ESTOI.

\section{Experiments}
\label{sec:sec4}
\subsection{Data Preparation}
\label{sec:sec4:sub1}
Speech materials consisted of Harvard sentences \cite{rothauser1969ieee} spoken by two (one male \cite{valentini2019hurricane} and one female \cite{demonte2019}) native English speakers. The Harvard sentences are organized as 72 sets of 10 sentences each, and each set is designed to be phonetically balanced. Sentences were selected from sets 1--60, 61--66, and 67--72 for training, validation, and test data, respectively. 

Six types of background noise were used: babble, restaurant, station, cafeteria, airport announcement, and speech-shaped noise (SSN), with the first five from the MS-SNSD dataset \cite{reddy2019scalable} and SSN artificially generated by VOICEBOX \cite{brookes2005voicebox}. For training and validation data, we selected four types of noise (babble, station, restaurant, and SSN) to generate noisy speech at three SNR levels, i.e., --11, --7, and --3 dB. 
The remaining two types of noise were used for test data. For cafeteria noise, the SNRs were set to --9, --5, and --1 dB; for airport announcements noise, they were set to --13, --9, and --5 dB.

Although reverberation was disregarded in the training, we examined if the proposed system can work well in reverberant environments. Besides the original room condition (recorded in professional studios with reverberation time $T_{60}\approx0.30$~s), another two RIRs were selected from a large room ($T_{60}=0.61$~s) in the MIRD database \cite{hadad2014multichannel} and stairway ($T_{60}=0.92$~s) in the AIR database \cite{jeub2009binaural}. Thus, there were a total of three (1 original $+$ 2 selected RIRs) reverberant environments considered in the test set. When generating noisy-reverberant speech, we first convolved the raw speech with the RIR, and then added the masker noise to the obtained reverberant speech at a desired SNR level. 

To summarize, there were 14,400 (600 sentences $\times$ 2 genders $\times$ 3 SNRs $\times$ 4 noises) utterances in the training set; 1,440 (60 sentences $\times$ 2 genders $\times$ 3 SNRs $\times$ 4 noises) utterances in the validation set; and 2,160 (60 sentences $\times$ 2 genders $\times$ 3 SNRs $\times$ 2 noises $\times$ 3 reverberations) utterances in the test set. For the test set, a total of 18 listening conditions (comprising of 3 SNRs, 2 noises, and 3 reverberations) were extensively evaluated. It is worth noting that all the sentences, noises, reverberations (except the original condition), and SNR levels of the test set were unseen during model training.

\vspace{-1mm}
\subsection{Implementation Details}
\label{sec:sec4:sub2}
All signals were down-sampled to 16 kHz in our experiments. For feature extraction, we first used a Hanning window with a window size of 32 ms and hop size of 16 ms to compute the spectrogram. Next, 64 ERB-scaled triangular bands were applied to the spectrogram to produce the 64-dimensional input features for neural networks. All the input features were power-law compressed with a power of $1/6$.
We chose SSDRC \cite{ZorilaKS12} as the reference algorithm to generate the signal example $\hat{y}$ that was used in Equations~(\ref{eq:L_D_intel}) and (\ref{eq:L_D_qua}).
During training, we normalized all metric scores to the range of $[0, 1]$, i.e., the same range with sigmoid activation, and set the target maximum scores ($t_{\rm{int}}$ and $t_{\rm{qua}}$ in Equation~(\ref{eq:G_update})) to 1.
Specifically, we used the following parametric logistic function for score normalization:
\begin{equation}
    f(v) = \frac{1}{1+\exp(a*(v-b))},
    \label{eq:normalize_logistic}
\end{equation}
where $v$ denotes the raw metric score. Parameters $(a,b)$ were accordingly set as $(-0.06, 32)$ for SIIB; $(-0.95, 2.8)$ for HASPI; $(-8.0, 0.25)$ for ESTOI; $(-1.5, 2.5)$ for PESQ; and $(-2.5, 2.2)$ for ViSQOL. These parameters were empirically chosen to make the normalized scores uniformly distributed between 0 and 1, which helps reduce bias and stabilize GAN training.

For GAN model configurations, the Adam optimizer \cite{kingma2014adam} was used in the training, with initial learning rates of
0.0004 and 0.0002 for the generator ($G$) and the discriminators ($D_{\rm{int}}$ and $D_{\rm{qua}}$), respectively. The batch size was 1, and the hyper-parameter $\lambda$ in Equation~(\ref{eq:G_update}) was set to 0.5.
The training process was terminated when all three intelligibility scores (SIIB, HASPI, and ESTOI) on the validation set stopped improving for five consecutive epochs\footnote{Source codes and the pre-trained model are available at \url{https://github.com/nii-yamagishilab/NELE-GAN}}.

\begin{table*}[t]
    \caption{Average objective scores of the compared systems across different reverberant conditions under \textbf{cafeteria} noise.}
    \vspace{1.0mm}
    \label{tab:result_on_caf}
    \centering
    \renewcommand\tabcolsep{3.3pt}
    \renewcommand\arraystretch{1.45}
    \scalebox{0.999}{
    \begin{tabular}{p{61.4pt}<{\centering} p{22pt}<{\centering}p{24pt}<{\centering}p{23pt}<{\centering}p{24.5pt}<{\centering} c p{22pt}<{\centering}p{24pt}<{\centering}p{23pt}<{\centering}p{24.5pt}<{\centering} c p{22pt}<{\centering}p{24pt}<{\centering}p{23pt}<{\centering}p{24.5pt}<{\centering} c p{22pt}<{\centering}p{28.5pt}<{\centering}}
        \hline
        \hline
           \multirow{2}{*}{System} &
           \multicolumn{4}{c}{Intelligibility in $T_{60}\approx0.30$ s} & & \multicolumn{4}{c}{Intelligibility in $T_{60}=0.61$ s} & &
           \multicolumn{4}{c}{Intelligibility in $T_{60}=0.92$ s} & &
           \multicolumn{2}{c}{Quality} \\
           \cline{2-5}  \cline{7-10} \cline{12-15} \cline{17-18}
            & SIIB & HASPI & ESTOI & sEPSM & & SIIB & HASPI & ESTOI & sEPSM & & SIIB & HASPI & ESTOI & sEPSM & & PESQ & ViSQOL \\
           \hline
          Unmodified & 15.90 & 1.92 & 0.228 & 6.70 & & 15.76 & 1.77 & 0.220 & 6.61 & & 9.26 & 1.42 & 0.134 & 5.89 & & 4.50 & 5.00 \\
          \hline
          SSDRC & 30.98 & 2.74 & 0.314 & 7.03 & & 24.72 & 2.27 & 0.273 & 6.77 & & 15.24 & 1.83 & 0.199 & 6.04 & & 3.52 & \textbf{2.71} \\
          iMetricGAN & 35.61 & 2.85 & 0.302 & 7.16 & & 26.90 & 2.34 & 0.256 & 6.88 & & 16.44 & 1.89 & 0.193 & 6.14 & & 3.20 & 2.56 \\
          \hline
          S-GAN & 37.89 & 2.77 & 0.239 & 7.31 & & 30.57 & 2.35 & 0.208 & 7.04 & & 17.91 & 1.79 & 0.154 & 6.20 & & 2.08 & 2.02 \\
          H-GAN & 35.12 & \textbf{3.12} & 0.242 & \textbf{7.55} & & 27.58 & 2.61 & 0.205 & 7.13 & & 16.57 & 1.99 & 0.149 & 6.28 & & 2.07 & 2.08 \\
          E-GAN & 34.20 & 2.71 & \textbf{0.331} & 7.21 & & 28.17 & 2.36 & \textbf{0.285} & 6.94 & & 16.03 & 1.81 & 0.207 & 6.15 & & 3.07 & 2.38 \\
          Proposed (S+H+E) & \textbf{41.33} & 3.11 & 0.313 & 7.53 & & \textbf{32.99} & \textbf{2.62} & 0.268 & \textbf{7.17} & & \textbf{18.90} & \textbf{2.00} & 0.194 & \textbf{6.28} & & 2.63 & 2.17 \\
          Proposed (All) & 37.97 & 2.95 & 0.324 & 7.44 & & 31.05 & 2.52 & 0.277 & 7.11 & & 18.48 & 1.96 & \textbf{0.209} & 6.26 & & \textbf{3.54} & 2.69 \\
        \hline
        \hline
    \end{tabular}
    \vspace{2mm}
    }
\end{table*}

\begin{table*}[t]
    \caption{Average objective scores of the compared systems across different reverberant conditions under \textbf{airport announcement} noise.}
    \vspace{1.0mm}
    \label{tab:result_on_airport}
    \centering
    \renewcommand\tabcolsep{3.3pt}
    \renewcommand\arraystretch{1.45}
    \scalebox{0.999}{
    \begin{tabular}{p{61.4pt}<{\centering} p{22pt}<{\centering}p{24pt}<{\centering}p{23pt}<{\centering}p{24.5pt}<{\centering} c p{22pt}<{\centering}p{24pt}<{\centering}p{23pt}<{\centering}p{24.5pt}<{\centering} c p{22pt}<{\centering}p{24pt}<{\centering}p{23pt}<{\centering}p{24.5pt}<{\centering} c p{22pt}<{\centering}p{28.5pt}<{\centering}}
        \hline
        \hline
           \multirow{2}{*}{System} &
           \multicolumn{4}{c}{Intelligibility in $T_{60}\approx0.30$ s} & & \multicolumn{4}{c}{Intelligibility in $T_{60}=0.61$ s} & &
           \multicolumn{4}{c}{Intelligibility in $T_{60}=0.92$ s} & &
           \multicolumn{2}{c}{Quality} \\
           \cline{2-5}  \cline{7-10} \cline{12-15} \cline{17-18}
            & SIIB & HASPI & ESTOI & sEPSM & & SIIB & HASPI & ESTOI & sEPSM & & SIIB & HASPI & ESTOI & sEPSM & & PESQ & ViSQOL \\
          \hline
          Unmodified & 16.25 & 2.20 & 0.191 & 6.63 & & 16.12 & 2.07 & 0.190 & 6.61 & & 9.43 & 1.58 & 0.115 & 5.79 & & 4.50 & 5.00 \\
          \hline
          SSDRC & 32.49 & 3.38 & 0.286 & 7.24 & & 25.80 & 2.71 & 0.261 & 6.85 & & 16.37 & 2.17 & 0.203 & 6.06 & & 3.52 & \textbf{2.71} \\
          iMetricGAN & 35.68 & 3.44 & 0.280 & 7.37 & & 27.72 & 2.73 & 0.250 & 6.95 & & 17.98 & 2.23 & 0.204 & 6.18 & & 3.22 & 2.58 \\
          \hline
          S-GAN & 42.34 & 3.54 & 0.214 & 7.82 & & 34.21 & 2.85 & 0.195 & 7.26 & & 21.75 & 2.25 & 0.160 & 6.30 & & 2.12 & 2.04 \\
          H-GAN & 39.19 & \textbf{3.80} & 0.226 & 7.89 & & 31.50 & 3.03 & 0.201 & 7.34 & & 20.25 & \textbf{2.41} & 0.165 & 6.37 & & 2.08 & 2.10 \\
          E-GAN & 35.04 & 3.36 & 0.283 & 7.39 & & 28.88 & 2.82 & \textbf{0.263} & 7.03 & & 18.09 & 2.23 & 0.205 & 6.17 & & 3.07 & 2.40 \\
          Proposed (S+H+E) & \textbf{43.45} & 3.75 & 0.279 & \textbf{7.94} & & \textbf{35.31} & \textbf{3.04} & 0.250 & \textbf{7.36} & & \textbf{22.36} & 2.40 & 0.206 & \textbf{6.37} & & 2.71 & 2.19 \\
          Proposed (All) & 42.54 & 3.72 & \textbf{0.288} & 7.87 & & 34.30 & 3.00 & 0.257 & 7.30 & & 22.03 & 2.38 & \textbf{0.209} & 6.36 & & \textbf{3.56} & 2.67 \\
        \hline
        \hline
    \end{tabular}
    }
    \vspace{-0.5mm}
\end{table*}

\vspace{-1mm}
\subsection{Objective Evaluations}
\label{sec:sec4:sub3}
In this section, we evaluated the proposed system through objective measurements. We first re-implemented several baseline systems, and then conducted an ablation test, yielding a total of eight systems evaluated in the experiments. We explain and notate each system as follows:
\begin{itemize}
    \item \textbf{Unmodified:} Plain speech without any modification.
    \item \textbf{SSDRC:} A baseline system using the state-of-the-art SSDRC \cite{ZorilaKS12} algorithm, which achieved the highest and second highest intelligibility gains in the 1st \cite{cooke2013intelligibility} and 2nd \cite{Rennies2020} Hurricane challenges, respectively. It consists of two cascading non-parametric modifications: spectral shaping (SS) in frequency and dynamic range compression (DRC) in time.
    \item \textbf{iMetricGAN:} Our previously proposed system \cite{Li2020}, in which we used BLSTM networks to optimize SIIB and ESTOI. We re-implemented it with the same model configurations, except the original sampling rate (44.1 kHz) was adjusted to 16 kHz. Its model size was 7.8M parameters, which is much larger than the proposed system (2.1M parameters for $G$).
    \item \textbf{S-GAN:} A system optimizing only SIIB, in which $D_{\rm{int}}$ was simplified to predict only a single SIIB score, and no $D_{\rm{qua}}$ was used for optimizing quality metrics.
    \item \textbf{H-GAN:} A system optimizing only HASPI.
    \item \textbf{E-GAN:} A system optimizing only ESTOI.
    \item \textbf{Proposed (S+H+E):} A partial version of our proposed system jointly optimizing three intelligibility metrics, i.e., SIIB, HASPI, and ESTOI. No $D_{\rm{qua}}$ was used for optimizing quality metrics.
    \item \textbf{Proposed (All):} Our full proposed system jointly optimizing three intelligibility metrics (SIIB, HASPI, and ESTOI) and two quality metrics (PESQ and ViSQOL).
\end{itemize}

\begin{figure*}[t]
    \centering
    \subfigure[Under cafeteria noise.]{
        \includegraphics[height=86.2mm,width=88.5mm]{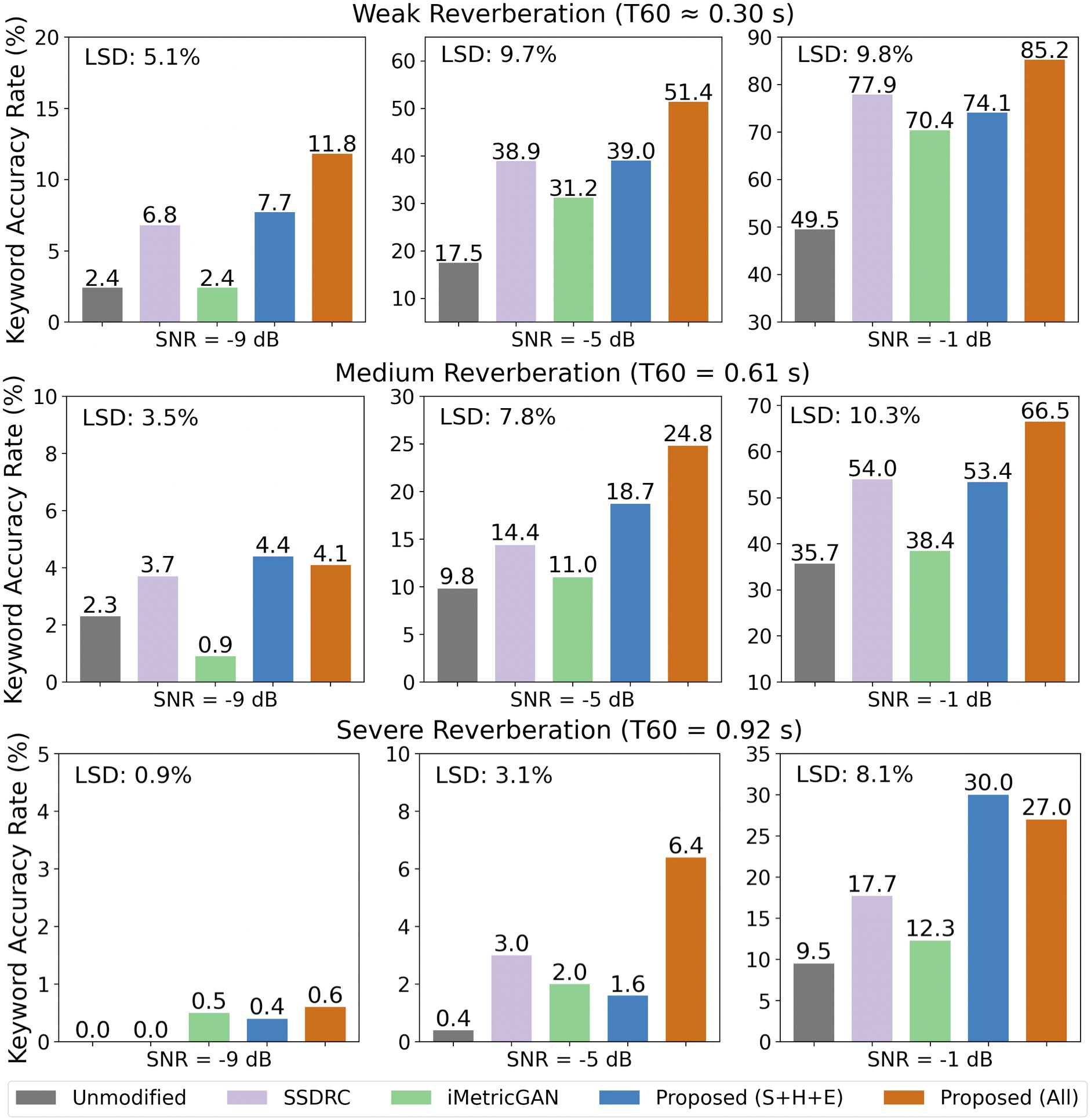}}
    \subfigure[Under airport announcement noise.]{
        \includegraphics[height=86.2mm,width=88.5mm]{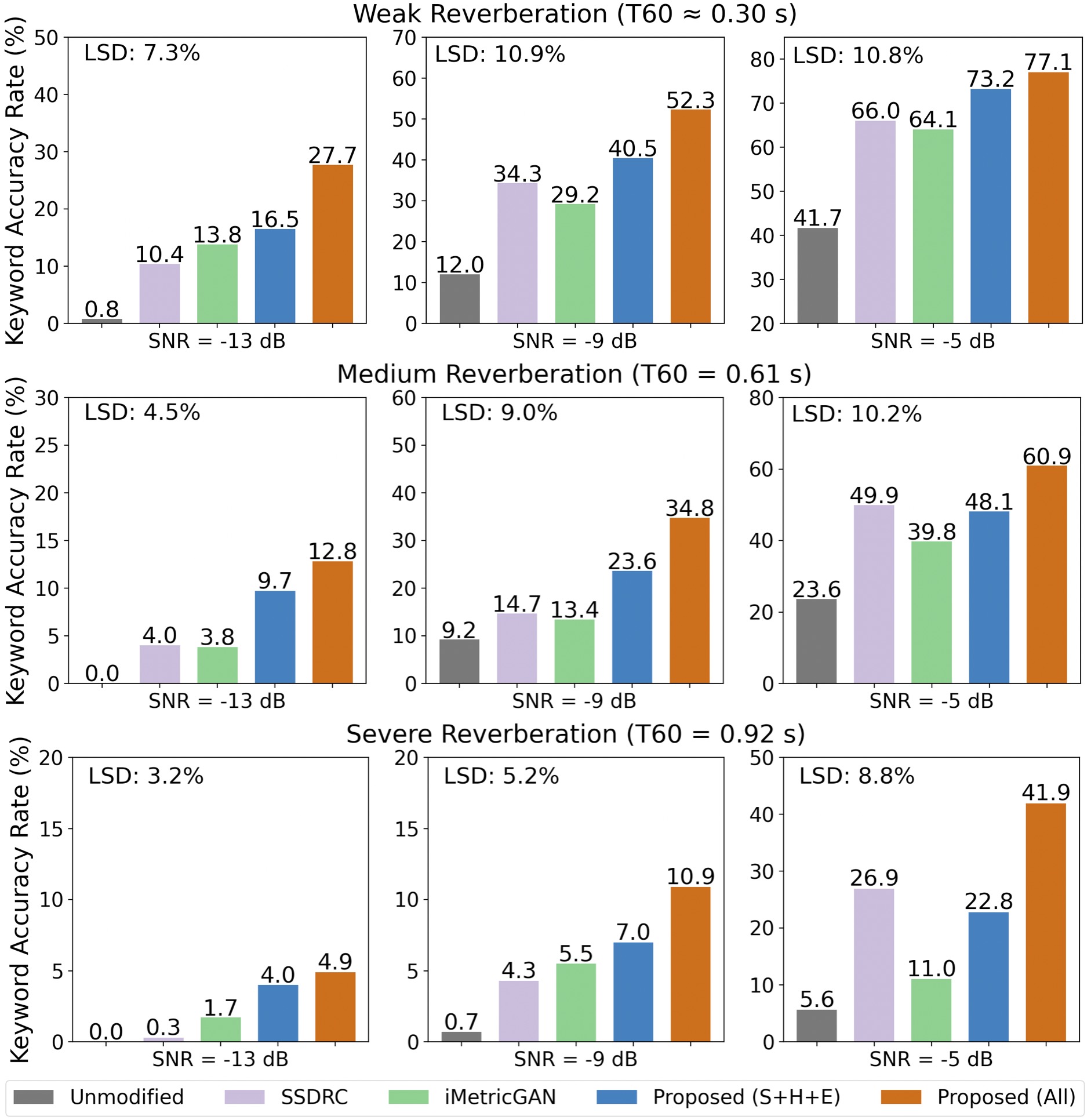}}
    \caption{Mean keyword accuracy rates (KARs) in percentage points for each compared system across different listening conditions.}
    \label{fig:KAR_result}
    \vspace{-2mm}
\end{figure*}

\begin{figure}[t]
    \centering
        \includegraphics[height=49mm,width=88mm]{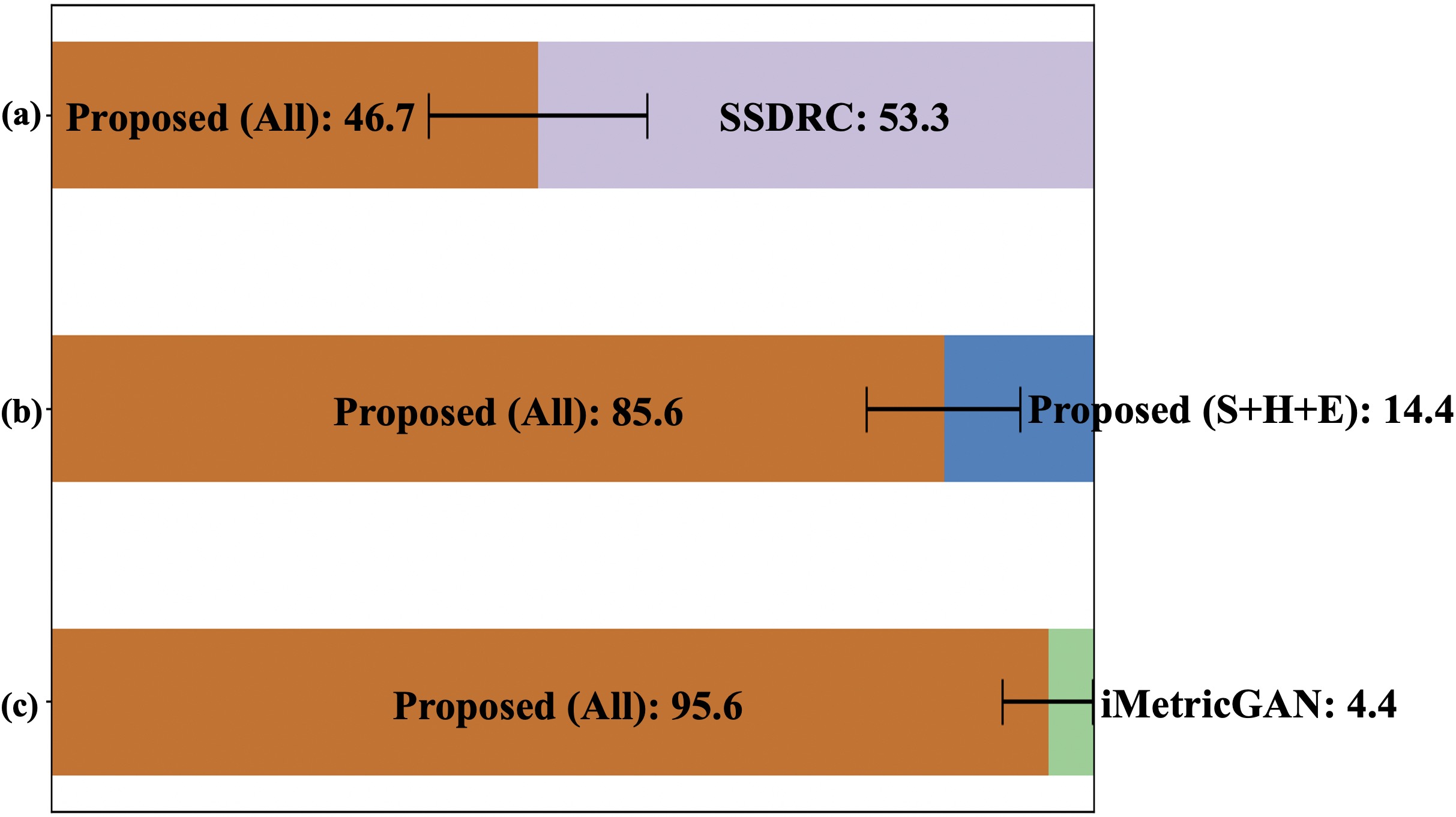}

    \caption{Preference scores (\%) with 95\% confidence intervals on speech quality compared between \mbox{\textbf{Proposed (All)}} and three reference systems.}
    \label{fig:preference}
    \vspace{-3mm}
\end{figure}

We used the same target metrics (SIIB, HASPI, and \mbox{ESTOI}) as the evaluation measurements due to their very high correlations with human perception \cite{van2018evaluation}.
Moreover, we incorporated an additional advanced metric called \mbox{sEPSM} \cite{steinmetzger2019predicting}, i.e., an improved intelligibility prediction model based on the speech envelope-power spectrum model \cite{jorgensen2013multi}.
Note that \mbox{sEPSM} was completely unseen to the model, it was thus regarded as a third-party evaluation measurement in the experiments, producing scores in the range of $[0, +\infty)$.
As discussed in Section~\ref{sec:sec4:sub1}, the objective intelligibility scores were extensively tested under two types of unseen noise under three room conditions: weak, medium, and severe reverberations\footnote{When computing the intelligibility scores under these reverberant conditions, the clean and distorted signals were time-aligned in advance.}. The quality scores (PESQ and ViSQOL) were computed by comparing the enhanced speech (without noise and reverberation) with input unmodified speech. For the above-mentioned six measurements, higher scores indicate better performance.

Tables~\ref{tab:result_on_caf} and \ref{tab:result_on_airport} list the average objective scores of each system under cafeteria and airport announcement noise, respectively. 
In both tables, \textbf{Proposed (All)} clearly outperformed the state-of-the-art baseline \textbf{SSDRC} in all room conditions with much higher intelligibility scores and comparable quality scores.
Benefiting from new target metrics and network architecture, it also consistently improved upon the previously proposed \textbf{iMetricGAN} for all six measurements with a far smaller model size.
Compared with \textbf{Proposed (S+H+E)}, \mbox{\textbf{Proposed (All)}} achieved much higher scores for speech quality with only a slight decrease in objective intelligibility scores\footnote{We also found that the quality scores can be further improved using a larger weight $\lambda$ in Equation~(\ref{eq:G_update}) at the cost of lower intelligibility scores.}.
\textbf{S-GAN}, \textbf{H-GAN}, and \textbf{E-GAN} performed well on their corresponding optimization targets.
For example, we can see that \textbf{H-GAN} achieved the best HASPI scores in some cases. However, there still remains quite a bit of room for improvement in terms of other non-target metrics. 
This indicates that optimizing only a single metric might cause sub-optimality in those unconsidered metrics.
By jointly optimizing multiple metrics, both \mbox{\textbf{Proposed (S+H+E)}} and \textbf{Proposed (All)} showed much more robust performance on all intelligibility measurements.
Specifically, \mbox{\textbf{Proposed (S+H+E)}} produced the best results in terms of unseen sEPSM scores, and this further demonstrates that the multi-metric optimization strategy can lead to effective and generalized intelligibility improvement.
More interestingly, we found that \mbox{\textbf{Proposed (S+H+E)}} and \mbox{\textbf{Proposed (All)}} can achieve extra SIIB gains even compared with the pure SIIB-oriented \mbox{\textbf{S-GAN}} system.

\subsection{Subjective Listening Tests}
\label{sec:sec4:sub4}

\begin{figure*}[t]
    \centering
    \subfigure[Unmodified]{
    \raisebox{-0.22mm}{
        \includegraphics[height=40.65mm,width=57.6mm]{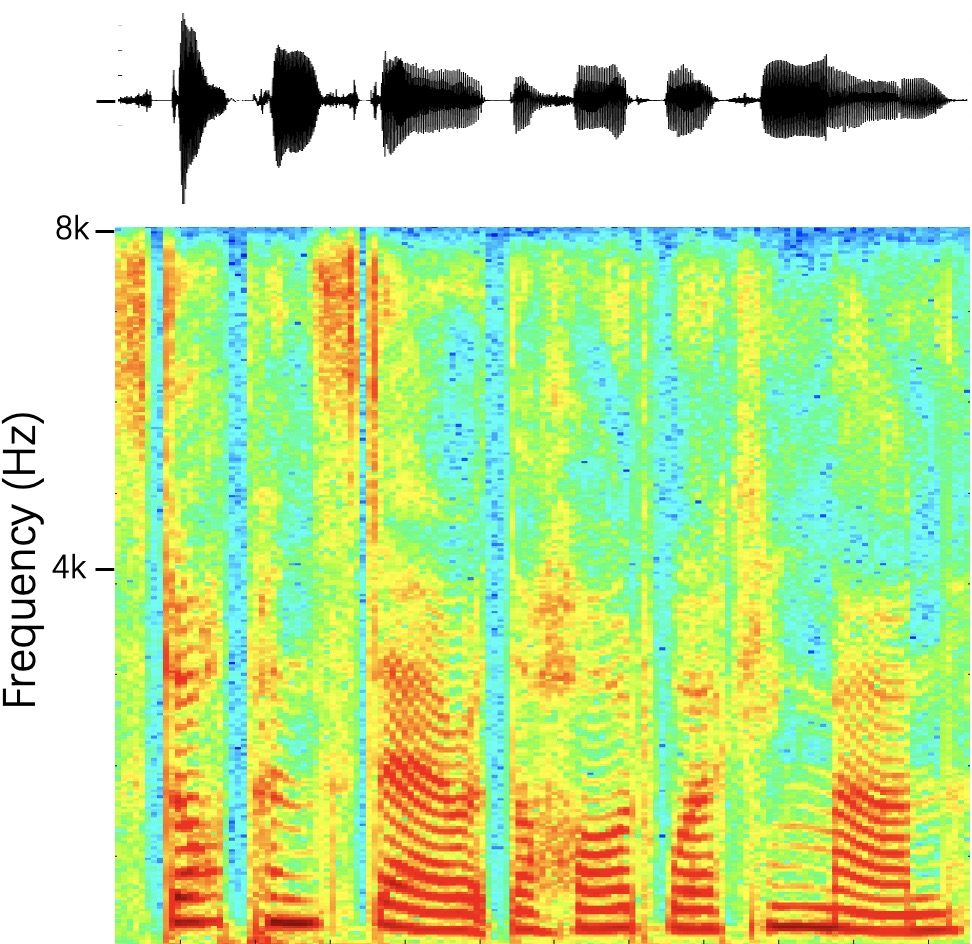}
        }
        }
    \subfigure[SSDRC]{
        \includegraphics[height=40mm,width=48mm]{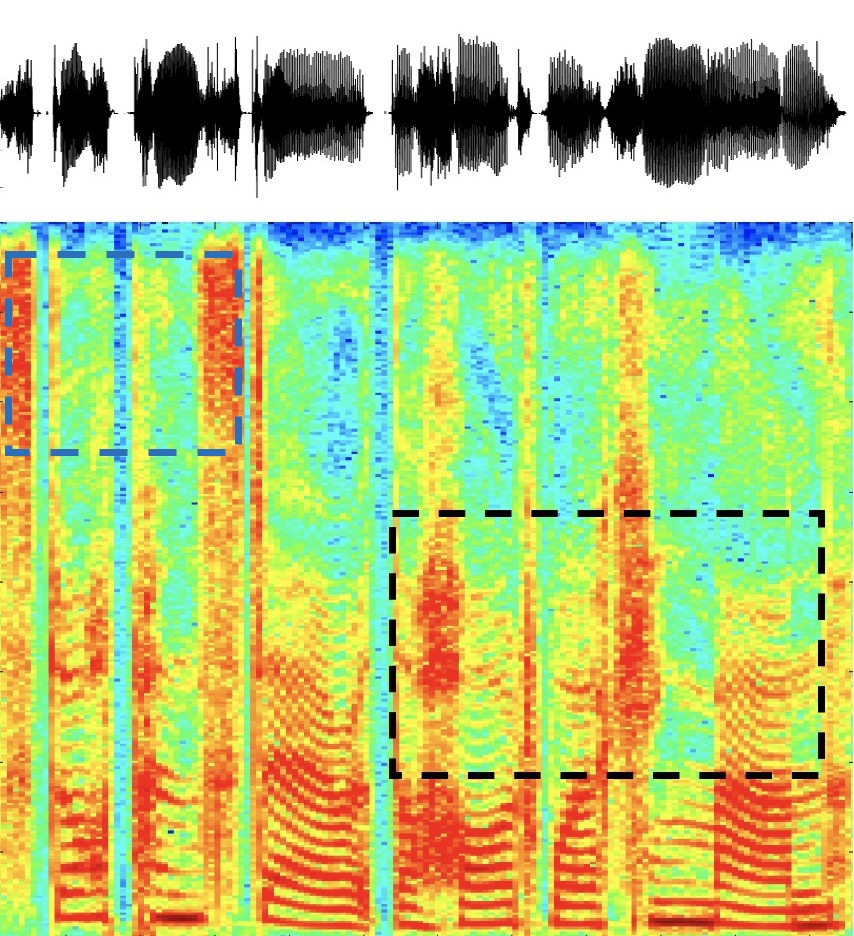}
        }
    \subfigure[Proposed (All)]{
        \includegraphics[height=40mm,width=48mm]{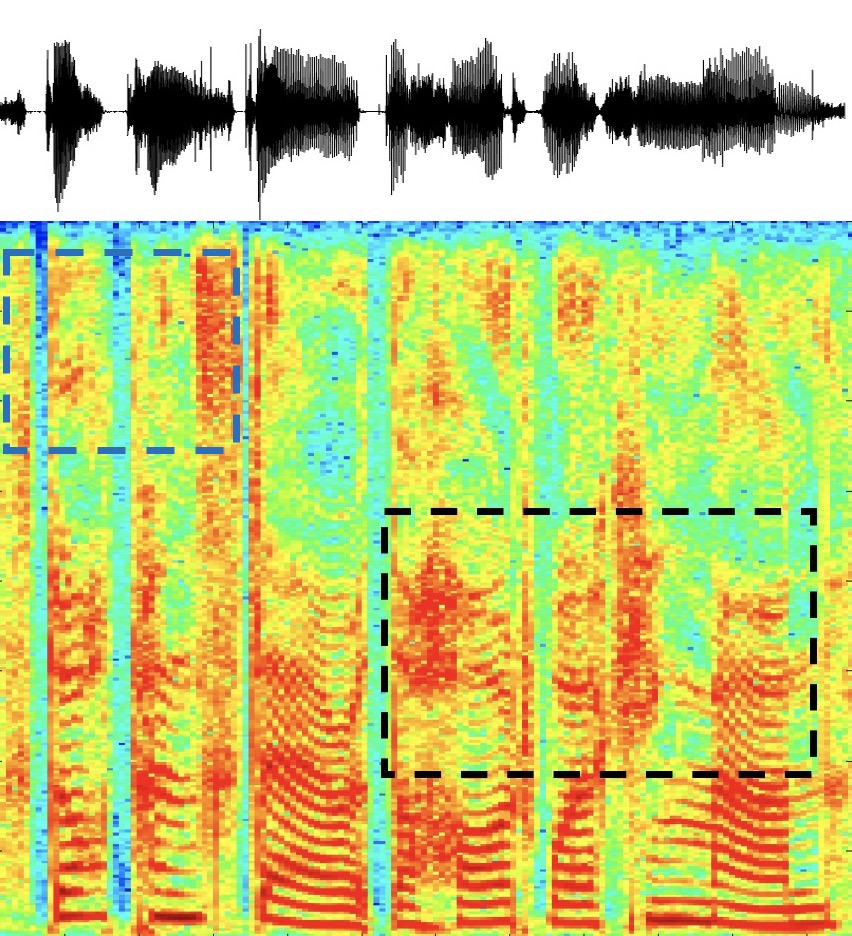}
        }
    \caption{Waveforms and their spectrograms on one utterance under cafeteria noise at SNR$=$--5 dB for different signals: (a) Unmodified input speech, (b) enhanced speech from \textbf{SSDRC}, and (c) enhanced speech from \textbf{Proposed (All)}. Utterance used is notated as ``f\_70\_8'', i.e., the 8-th utterance in 70-th list of female speaker.}
    \label{fig:examples_spectrogram}
    \vspace{-2mm}
\end{figure*}

\begin{figure*}[t]
    \centering
    \subfigure[LTAS gain under cafeteria noise at SNR$=$--5 dB]{
    \raisebox{0.1mm}{
        \includegraphics[height=42mm,width=88mm]{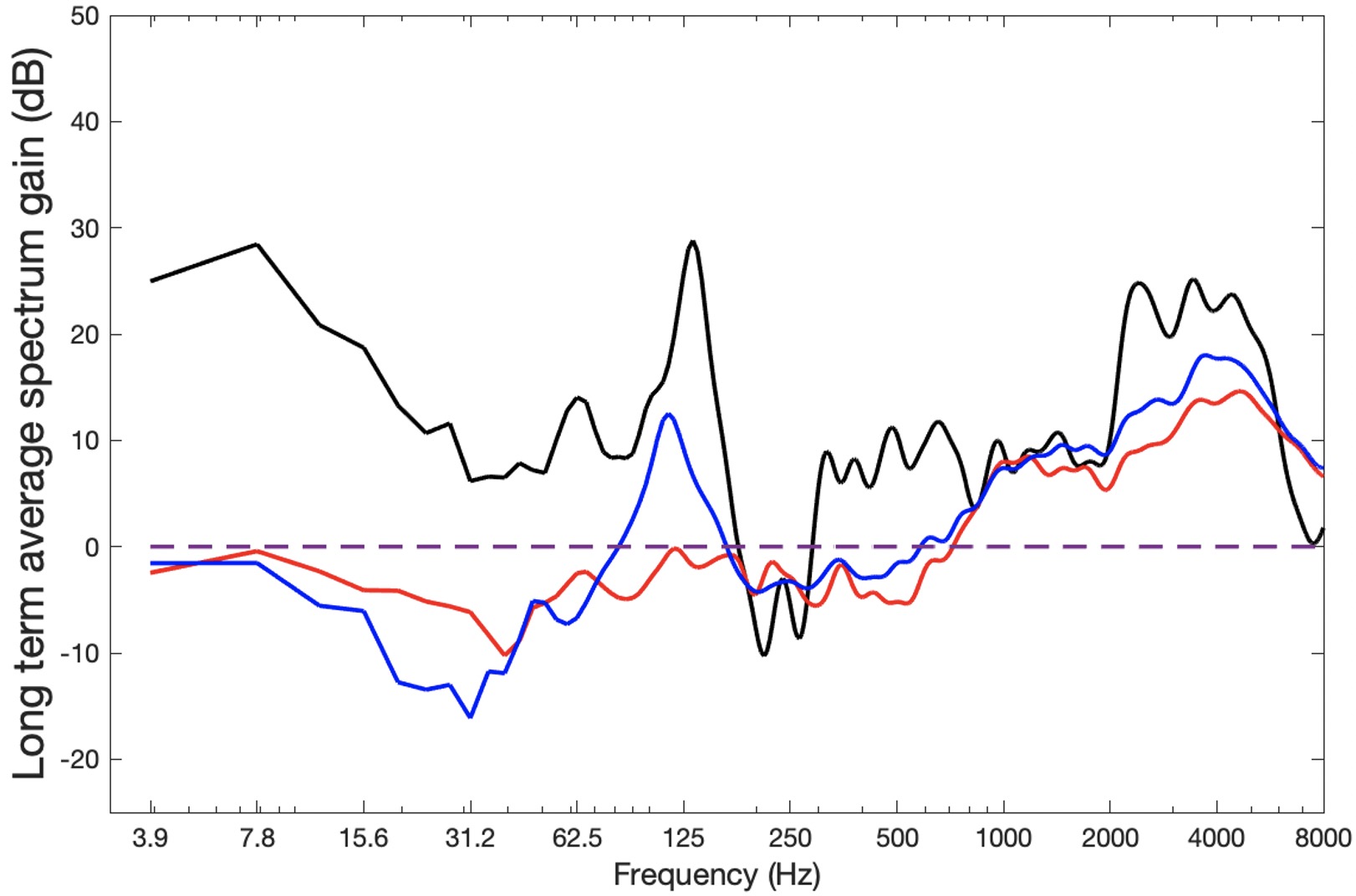}
        }
        }
    \subfigure[LTAS gain under airport announcement noise at SNR$=$--13 dB]{
        \includegraphics[height=42mm,width=84mm]{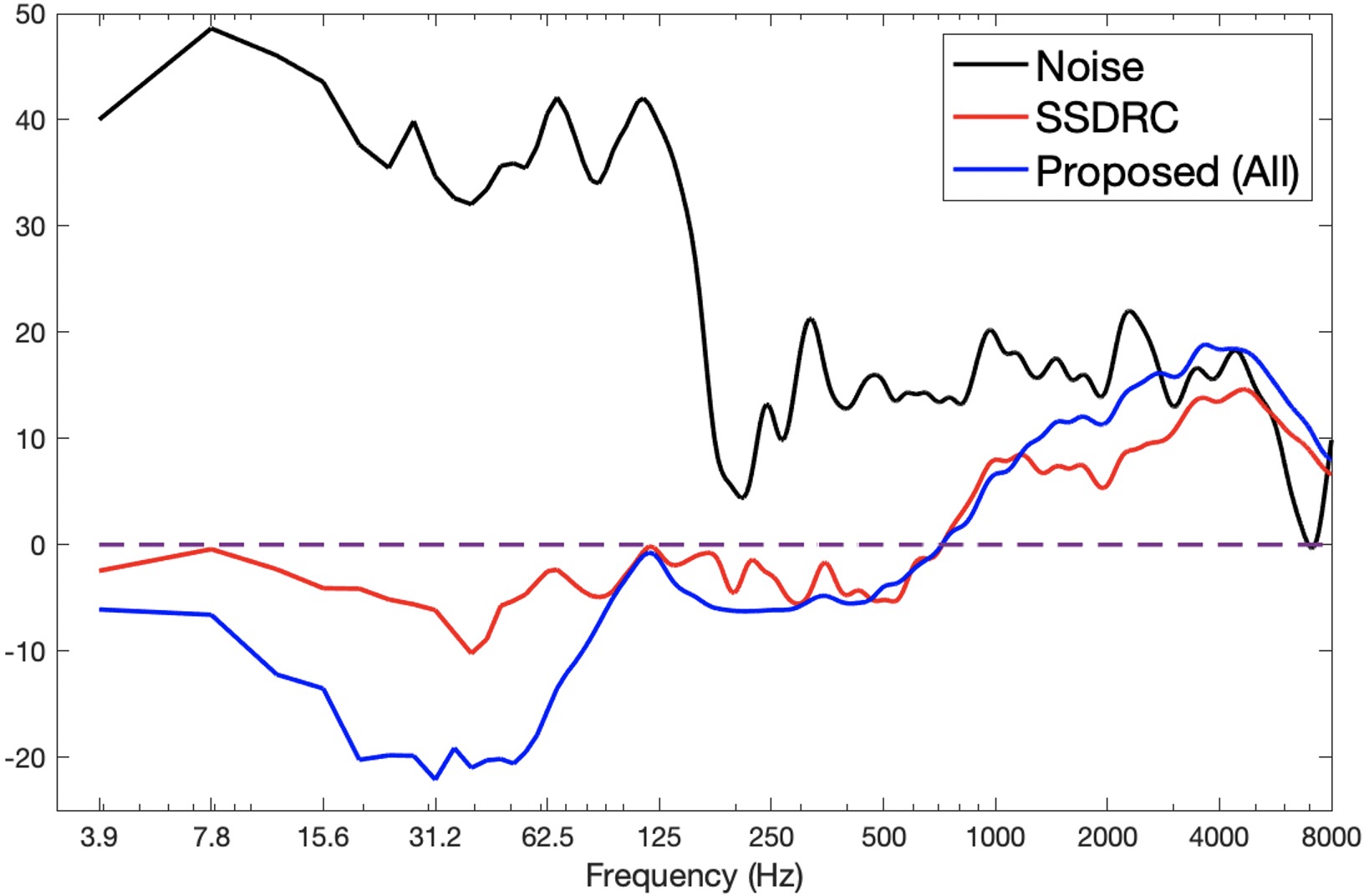}
        }

    \caption{Long-term average spectrum (LTAS) gain (dB) over LTAS of unmodified utterance (f\_70\_8) for: (1) masker noise, (2) enhanced speech from \textbf{SSDRC}, and (3) enhanced speech from \textbf{Proposed (All)}.}
    \label{fig:LTAS}
    \vspace{-2mm}
\end{figure*}

We conducted an intelligibility listening test to further evaluate the following five systems: \mbox{\textbf{Unmodified}}, \mbox{\textbf{SSDRC}}, \mbox{\textbf{iMetricGAN}}, \mbox{\textbf{Proposed (S+H+E)}}, and \mbox{\textbf{Proposed (All)}}.

60 Harvard sentences (sets 67, 69, and 71 of the female speaker; and sets 68, 70, and 72 of the male speaker) were extracted and presented in each of the 18 listening conditions (3 SNRs $\times$ 2 noises $\times$ 3 reverberations), producing a total of 5,400 tested utterances (60 sentences $\times$ 18 conditions $\times$ 5 systems). 
We then divided these tested utterances into 90 blocks: each block consisted of 60 individual Harvard sentences, and each Harvard sentence was processed using a random system and under a random listening condition. 
A total of 90 native English speakers with no reported hearing impairments were recruited for the online test, and all were paid. Each participant was assigned to one block. They were instructed to listen to each tested utterance only once then type in as many words they heard as possible. 
We also implemented a cheater-detection mechanism by assigning five additional validation utterances (with very slight noise) to each block of the main listening test. Participants who did not reach 60\% average word accuracy on these utterances were considered unqualified listeners, which led to three participants being excluded from the analysis. Following the evaluation rules of the 1st Hurricane challenge \cite{cooke2013intelligibility}, we only accounted for the correct content words in each transcription by excluding the short common words: `a', `the', `in', `to', `on', `is', `and', `of', and `for'.
The keyword accuracy rate (KAR) was then computed as the performance measure of intelligibility. 

The results are plotted in Fig.~\ref{fig:KAR_result}. Fisher's least significant difference (LSD) was also separately computed for each listening condition using \mbox{ANOVAs} to enable statistical comparisons of different systems. 
As shown in Fig.~\ref{fig:KAR_result}, modification algorithms can generally lead to substantial intelligibility gains to the unmodified speech, except for four extremely challenging conditions where all systems failed to reach 10\% KAR.
The best system in all but two of the 18 conditions was \mbox{\textbf{Proposed (All)}}. For all conditions, it consistently outperformed not only \mbox{\textbf{iMetricGAN}}, but also the state-of-the-art \mbox{\textbf{SSDRC}}. Interestingly, although its objective intelligibility scores were lower than those of \mbox{\textbf{Proposed (S+H+E)}} (see in Tables~\ref{tab:result_on_caf} and \ref{tab:result_on_airport}), it showed much higher increases in KAR. 
This reveals that incorporating quality metrics into training can largely contribute to subjective intelligibility, which is likely due to the effective suppression of audible artefacts\footnote{Audio samples of the tested systems are available at \url{https://nii-yamagishilab.github.io/hyli666-demos/intelligibility/index.html}}.

\begin{figure*}[t]
    \centering
        \includegraphics[height=49mm,width=181.5mm]{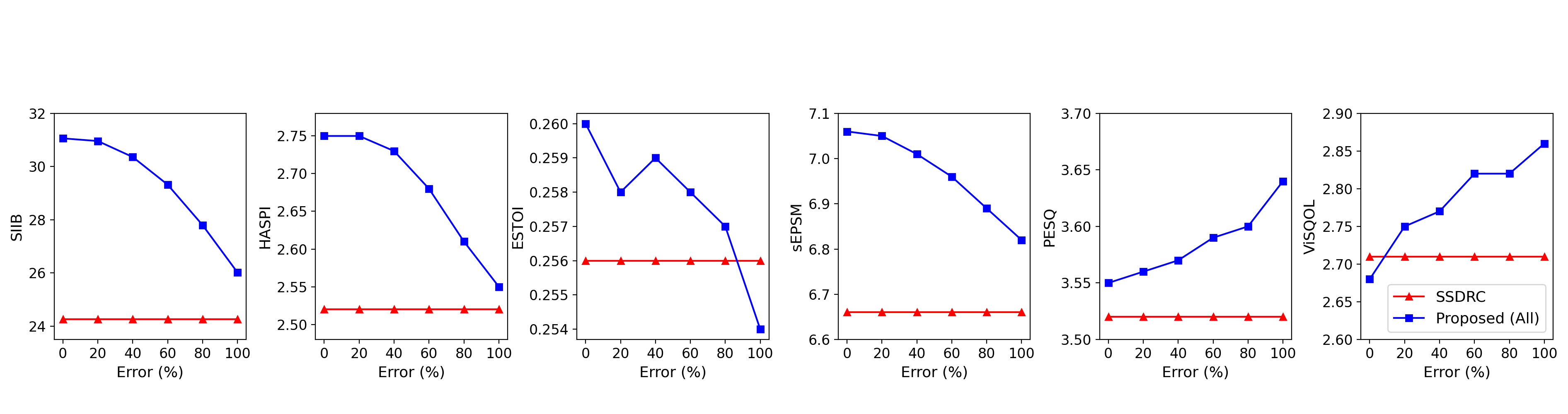}
         
    \caption{Average objective scores as noise estimation error is artificially added to noise PSD.}
    \label{fig:tolerance}
    \vspace{-2mm}
\end{figure*}

We also conducted AB preference tests to evaluate the perceptual quality of the enhanced speech. We conducted pairwise comparisons between \mbox{\textbf{Proposed (All)}} and the following three systems: (1) \mbox{\textbf{SSDRC}}; (2) \mbox{\textbf{iMetricGAN}}; and (3) \mbox{\textbf{Proposed (S+H+E)}}.
90 enhanced samples were randomly selected from the test set for each system, and a total of 15 listeners participated.
Each participant was instructed to listen to 18 randomized sample pairs, and for each pair they had to select the one that sounded better in terms of speech quality. 
As we can see from Fig.~\ref{fig:preference}, \mbox{\textbf{Proposed (All)}} achieved significantly higher preference scores than \mbox{\textbf{iMetricGAN}} and \mbox{\textbf{Proposed (S+H+E)}} and performed comparably with \textbf{SSDRC}.
Such results clearly indicate that speech quality can be effectively improved through incorporating objective quality metrics into model training. 
 
\vspace{-1mm}
\subsection{Acoustic Analysis on Enhanced Speech}
\label{sec:sec4:sub5}

We analyzed the acoustic properties of the enhanced speech. For deeper insight, we used \textbf{SSDRC} as the reference system to conduct a comparative study.
Figure~\ref{fig:examples_spectrogram} gives examples of waveforms and spectrograms for different signals. From the spectrograms, we found that both \textbf{SSDRC} and \textbf{Proposed (All)} modified the speech signal through redistributing its energy from low frequencies to the middle and high frequencies. By comparing Fig.~\ref{fig:examples_spectrogram}(c) with (b), \mbox{\textbf{Proposed (All)}} tended to allocate more energy on the middle-frequency regions (2$\sim$4 kHz) of the voiced segments (see black dashed box), while \textbf{SSDRC} emphasized the high-frequency regions (4$\sim$8 kHz) of the unvoiced segments (see blue dashed box). We can also see that the waveform envelope of the enhanced speech from \mbox{\textbf{Proposed (All)}} is similar to that of the original unmodified speech.
In contrast, the modified waveform of \mbox{\textbf{SSDRC}} drastically changed, resulting in more acoustic artefacts. 

We also investigated the gain (in dB) of the long-term average spectrum (LTAS) calculated over one unmodified utterance. 
The gain values indicate the energy level of a signal in a certain frequency region: the signal energy is higher than the unmodified utterance with gain $>$ 0 dB; otherwise, lower.
As shown in Fig.~\ref{fig:LTAS}, frequency regions from 1 kHz to 8 kHz were effectively boosted in both \textbf{SSDRC} and \textbf{Proposed (All)}, which accords with our observations in Fig.~\ref{fig:examples_spectrogram}. Different from noise-independent \textbf{SSDRC}, \textbf{Proposed (All)} can adapt well to the changing environments. For example, the noise in Fig.~\ref{fig:LTAS}(b) was extremely strong (up to 40 dB gain) in the low-frequency regions ($\sim$125 Hz). Thus, the system automatically gave up much more speech components in these regions, compared with how it performed under weaker noise in Fig.~\ref{fig:LTAS}(a).
We also found that the speech components between 65 Hz to 150 Hz were particularly boosted under cafeteria noise, as shown in Fig.~\ref{fig:LTAS}(a). Interestingly, this coincides with the properties of the cafeteria noise where a peak gain was also exhibited near the same regions (see blue and black lines).
We hypothesize that by increasing the speech components in such narrow but noise-dominant regions, the target speech can be differentiated from the surrounding noise in an easier manner through achieving a certain perception threshold. On the other hand, \textbf{SSDRC} performed merely the same processing of the speech with two different noises (see red lines); therefore, it cannot make full use of additional noise information. This is one of the points explaining why our proposed system performed better in both objective and subjective evaluations.

\subsection{Analysis of System Robustness}
\label{sec:sec4:robustness}
We further analyzed the system's robustness in two particular situations, where (1) speaker and language are unseen to the model; and (2) background noise estimation is not accurate.

\subsubsection{Speaker and language generalization}
\label{sec:spk_lang_gen}
We tested the proposed system on a separately-created German speaker test set to examine if it can work under the mismatched speaker and language conditions.
Specifically, we extracted 100 clean utterances from an unseen male German speaker \cite{Rennies2020} and set the same 18 listening conditions (i.e., 2 noise types, 3 SNRs and 3 room conditions) as used in the original test set (see Section~\ref{sec:sec4:sub1}), resulting in a total of 1,800 tested utterances. 
Table~\ref{tab:result_on_German} lists the objective evaluation results on this new German speaker test set, where the scores were averaged over all listening conditions. We can see that even though \textbf{\mbox{Proposed (All)}} was built only upon English training data, it still achieved significant intelligibility gains and outperformed \textbf{SSDRC} by a large margin. This further demonstrates that the proposed system is robust, which can generalize well to mismatched speaker and language.

\begin{table}[t]
    \caption{Average objective scores on new German speaker test set.}
    \label{tab:result_on_German}
    \centering
    \renewcommand\tabcolsep{1.5pt}
    \renewcommand\arraystretch{1.35}
    \scalebox{1.115}{
    \begin{tabular}{p{48pt}<{\centering} c
    p{22pt}<{\centering} p{24pt}<{\centering} p{24pt}<{\centering} p{25pt}<{\centering} c
    p{22pt}<{\centering}p{28.5pt}<{\centering}}
        \hline
        \hline
           \multirow{2}{*}{System} &
           &
           \multicolumn{4}{c}{Intelligibility} & & 
           \multicolumn{2}{c}{Quality} \\
           \cline{3-6} \cline{8-9}
            & & SIIB & HASPI & ESTOI & sEPSM & & PESQ & ViSQOL \\
           \hline
          Unmodified & & 12.64 & 1.63 & 0.167 & 6.65 & & 4.50 & 5.00 \\
          \hline
          SSDRC  & & 25.27 & 2.40 & 0.252 & 7.00 & & 3.40 & 2.58 \\
          Proposed (All) & & \textbf{28.94} & \textbf{2.66} & \textbf{0.254} & \textbf{7.44} & & \textbf{3.46} & \textbf{2.81} \\
        \hline
        \hline
    \end{tabular}
    }
    \vspace{-1.5mm}
\end{table}

\subsubsection{Tolerance to noise estimation error}
\label{sec:tolerance_to_noise}
Next, we measured the tolerance of the proposed system to inaccuracy of background noise estimation. As described in Section~\ref{sec:sec2}, in order to exploit noise information, our system requires a reference microphone and runs IMCRA \cite{cohen2003noise} algorithm to estimate noise PSD, i.e., $W^2(m,k)$. However, such noise estimation might be inaccurate, for example, when the noise is highly non-stationary or the reference microphone is distant from the listener's position. To simulate estimation error in this process, we randomly marked certain noise PSD bins as error bins with an error rate of $\epsilon\%$; thus, the corrupted noise PSD $ W_{e}^2(m,k)$ is given as follows:
\begin{equation}
\label{eq:error}
    W_{e}^2(m,k) = 
\begin{cases}
exp(N), & if \hspace{0.15cm} error \\
W^2(m,k), & else
\end{cases}
\end{equation}
where $N$ is the random noise generated from Gaussian distribution with the same mean and variance as those of $logW^2(m,k)$, and error rate controls the corruption level: a higher $\epsilon\%$ indicates that each estimated bin is more likely filled with random noise, making noise estimation more inaccurate. 

\begin{table*}[t]
    \caption{Average objective scores for systems with different normalization methods on test set.}
    \label{tab:result_on_normalization}
    \centering
    \renewcommand\tabcolsep{5.2pt}
    \renewcommand\arraystretch{1.4}
    \scalebox{1.12}{
    \begin{tabular}{p{47pt}<{\centering}
    p{32pt}<{\centering} p{41.8pt}<{\centering} c
    p{25pt}<{\centering}p{25pt}<{\centering}p{25pt}<{\centering}p{25pt}<{\centering} c
    p{25pt}<{\centering}p{28.5pt}<{\centering}}
        \hline
        \hline
           \multirow{2}{*}{\makecell[c]{Normalization\\method}} &
           \multirow{2}{*}{Causal} &
           \multirow{2}{*}{\makecell[c]{Equal-power\\constrained}} &
           &
           \multicolumn{4}{c}{Intelligibility} & & 
           \multicolumn{2}{c}{Quality} \\
           \cline{5-8} \cline{10-11}
            & & & & SIIB & HASPI & ESTOI & sEPSM & & PESQ & ViSQOL \\
           \hline
          Unmodified & -- & -- & & 13.79 & 1.82 & 0.180 & 6.37 & & 4.50 & 5.00 \\
          \hline
          P-All-UL & \scalebox{0.80}{\XSolid} & \Checkmark & & 31.06 & 2.75 & 0.260 & 7.06 & & 3.55 & 2.68 \\
          P-All-FL & \Checkmark & \Checkmark & & 20.15 & 2.26 & 0.193 & 6.73 & & 3.29 & 2.53 \\
          P-All-Soft & \Checkmark & \scalebox{0.80}{\XSolid} & & 29.79 & 2.68 & 0.249 & 7.06 & & 3.55 & 2.67 \\
        \hline
        \hline
    \end{tabular}
    }
\end{table*}

Figure~\ref{fig:tolerance} shows the objective metric scores under different error rates. 
For the intelligibility metrics (i.e., SIIB, HASPI, ESTOI, and sEPSM), the corrupted noise PSD did not affect performance much when the error rate $\epsilon$ was less than 40\% but decreased intelligibility scores incrementally when $\epsilon > 40\%$.
However, even when noise estimation completely failed (i.e., $\epsilon=100\%$), \textbf{\mbox{Proposed (All)}} could still surpass the performance of \textbf{SSDRC} in intelligibility metrics (except ESTOI), which demonstrates that the proposed system is very robust against noise estimation error. 
From another point of view, by simply substituting random values for noise PSD, \textbf{\mbox{Proposed (All)}} degenerates into a noise-independent system. This also indicates that our system is flexible and can adapt to scenarios in which the implementation of a reference microphone is not available.
More interestingly, we found that the quality metrics (PESQ and ViSQOL) increased with increasing noise estimation error. We hypothesize that the system tends to modify the speech in a relatively aggressive manner to fully make use of noise information, e.g., giving up much more speech components in low-frequency regions when low-frequency noise is strong (see blue line in Fig.~\ref{fig:LTAS}(b)). For larger $\epsilon\%$, the system cannot exploit useful information as the given noise PSD becomes random; therefore, it tends to perform moderate modification, resulting in higher quality scores.

\vspace{-1mm}
\subsection{Extensions to Real-Time Execution}
\label{sec:sec4:realtime}

\begin{figure}[t]
    \centering
    \subfigure[Histogram.]{
    \raisebox{-1.9mm}{
        \includegraphics[height=50mm,width=48mm]{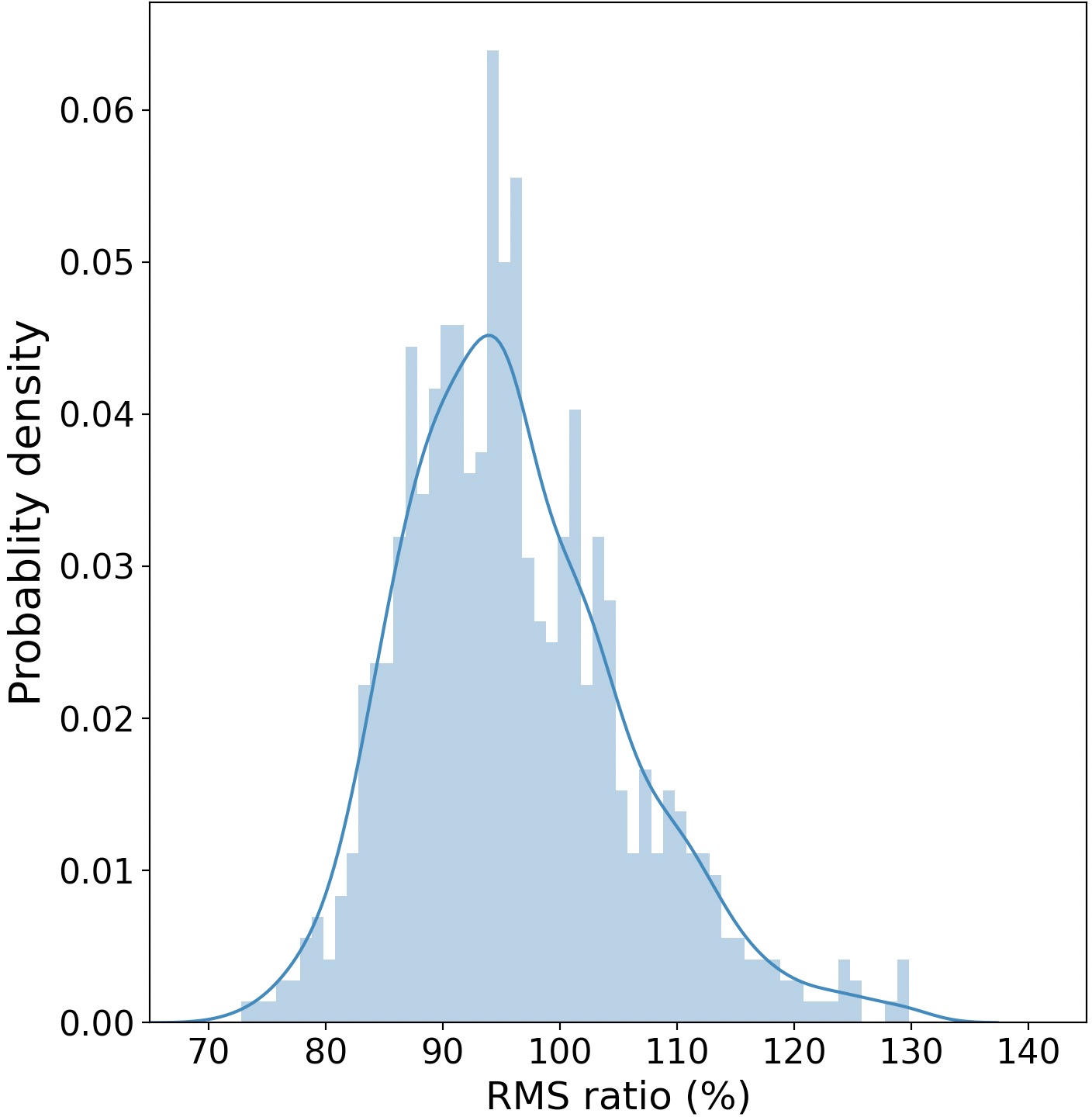}
        }
        }
    \subfigure[Box plot.]{
        \includegraphics[height=48.5mm,width=28mm]{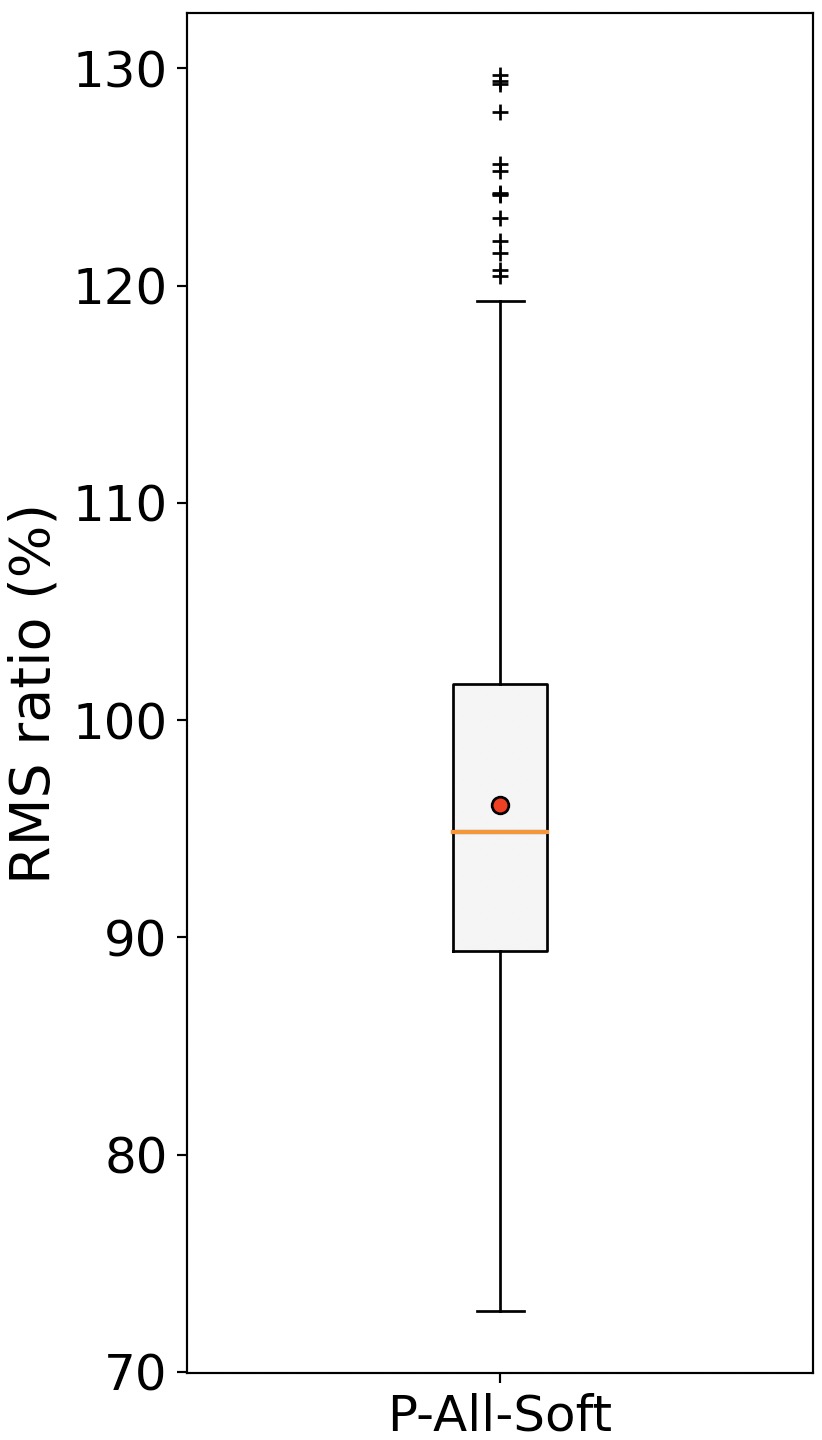}
        }

    \caption{Statistical results ($\gamma$ used in \textbf{P-All-Soft} was set to 5.62) of RMS ratios between enhanced and unmodified raw speech: (a) frequency density histogram of RMS ratios, and (b) box plot on RMS ratios, with red dot representing mean score.}
    \label{fig:soft_energy}
    \vspace{-2mm}
\end{figure}

Real-time execution is crucial for many speech applications such as mobile telephony. In this section, we discuss the causality of the proposed system in detail.
As discussed in Section~\ref{sec3:sub3:sub1}, the $G$ used in \mbox{\textbf{Proposed (All)}} can inherently perform intelligibility enhancement at the frame level in a causal manner. However, due to the equal-power constraint of Equation~(\ref{eq:power_constraint}), we still need to collect the entire signal to calculate the global energy of an utterance. Thus, we consider two extended methods to overcome this limitation. 

First, we revised the original utterance-level normalization (Equation~(\ref{eq:power_constraint})) to the following frame-level normalization:
\begin{equation}
    \label{eq:power_constraint_framelevel}
    \sum\limits_{i}\alpha^2(m,i)E_x(m,i)=\sum\limits_{i}E_x(m,i), \quad \forall m.
\end{equation}
As shown in Equation~(\ref{eq:power_constraint_framelevel}), the energy is normalized at each frame $m$ instead of the whole utterance, which enables the system to perform real-time execution under the equal-power constraint. We denote this modified frame-level normalization method for our proposed system as \textbf{P-All-FL}. Compared with the original proposed system with the utterance-level normalization method (denoted as \textbf{P-All-UL}), \textbf{P-All-FL} can only redistribute the speech energy across the frequency bands within one frame but not perform inter-frame redistribution.

Second, we consider another normalization method for application scenarios in which the equal-power constraint is not rigorous. 
As mentioned in Section~\ref{sec3:sub3:sub1}, the global scale factor $\gamma$ originally used in \mbox{\textbf{P-All-UL}} is calculated dynamically for each utterance to achieve perfect energy normalization.
With this method, however, we prepare such a $\gamma$ in advance by statically calculating the average energy ratio between the unmodified and enhanced speech over the whole training set.
The $\gamma$ is then applied to the raw amplification factors to compensate for the energy loss, achieving a soft energy normalization where the enhanced speech has approximately the same energy with the unmodified one.
We denote this method as \textbf{P-All-Soft}, and $\gamma$ was calculated as 5.62 from the training set.
Figure~\ref{fig:soft_energy} presents the statistical results of the root-mean-square (RMS) ratios between the enhanced and unmodified speech on the test set.
As shown in Fig.~\ref{fig:soft_energy}, the distribution of RMS ratios was concentrated close to one with a very small deviation. 
This indicates that the energy of enhanced speech can be well maintained within the approximately same level as the unmodified one by using \mbox{\textbf{P-All-Soft}} method.

Table~\ref{tab:result_on_normalization} lists the objective evaluation results on the three normalization methods. The scores were averaged over the whole test set across three SNR levels, three room conditions, and two unseen noises.
We found that \mbox{\textbf{P-All-FL}} did provide intelligibility gains to the unmodified speech. However, it performed much worse than the other two methods due to the lack of inter-frame energy distribution, which further reveals that energy reallocation in time is crucial for intelligibility improvement. 
Although \mbox{\textbf{P-All-Soft}} cannot perfectly fulfill the equal-power constraint, it satisfies the causality requirement and showed a comparable performance to \mbox{\textbf{P-All-UL}}. 
Note that all three methods differed only in the energy normalization strategy, while the core model of $G$ used in the experiments was identical.
By choosing a suitable normalization method in accordance with actual needs, the proposed system can satisfy different requirements of causality and energy constraint.

Finally, we give a brief analysis on the system complexity. The enhancement module, i.e., $G$, is composed of 2.1M weight parameters. Since each weight is used once for one multiply-add operation per frame (16 ms), $G$ thus takes 262.5 million floating-point operations per second (MFLOPS) for real-time execution\footnote{One multiply-add operation is counted as two operations.}. For other main modules, including two FFTs (for input speech and background noise analyses, respectively), one inverse FFT (for enhanced speech reconstruction), and IMCRA noise estimation, they take around 4.0 MFLOPS. The total complexity of the proposed system is around 270 MFLOPS. 
Considering both model size and the computational complexity, our proposed system is light-weight and can be easily implemented in practice.

\section{Conclusion}
\label{sec:sec5}

We proposed a GAN-based system for near-end speech intelligibility enhancement. 
To generate the intelligible and high-quality speech, we introduced a GAN model into our system to jointly optimize multiple intelligibility and quality metrics. 
Three modules are used in the GAN model to carry out such multi-metric optimization: an intelligibility discriminator that learns to predict the objective intelligibility scores of speech as accurately as possible, quality discriminator that similarly learns to predict the objective quality scores, and a generator that enhances the input speech signal to maximize both intelligibility and quality scores, which are computed with the above two discriminators, respectively.

Experimental results from both objective measurements and large-scale listening tests indicated that the proposed system can lead to significant intelligibility gains and perform much better than several compared baselines. 
It also generalizes well to various listening environments including unseen noises and reverberations. 
Moreover, the system is light-weight with only 2.1M parameters and can be easily extended to enable real-time execution.

\bibliography{main.bbl}

\begin{IEEEbiography}{Haoyu Li}
received the B.Sc.\ degree in physics from Nanjing University, China, in 2016, and the M.Eng.\ degree in electronic engineering from The University of Tokyo, Japan, in 2018.
He is currently a Ph.D.\ student with the SOKENDAI/National Institute of Informatics, Japan. His current research interests include statistical machine learning and speech enhancement.
\end{IEEEbiography}

\begin{IEEEbiography}{Junichi Yamagishi} (SM’13) received the Ph.D.\ degree from the Tokyo Institute of Technology (Tokyo Tech), Tokyo, Japan, in 2006. From 2007-2013 he was a research fellow in the Centre for Speech Technology Research (CSTR) at the University of Edinburgh, UK. He was appointed Associate Professor at National Institute of Informatics, Japan in 2013. He is currently a Professor at NII, Japan. He is also a Honorary Professor at the University of Edinburgh, UK. Since  2000,  he  has  authored and  co-authored  over  300  refereed  papers  in international  journals  and  conferences.  He  was  awarded  the  Itakura Prize from the Acoustic Society of Japan, the Kiyasu Special Industrial Achievement Award from the Information Processing Society of Japan, and the Young Scientists’ Prize from the Minister of Education, Science and Technology, the JSPS prize, the Docomo mobile science award in 2010, 2013, 2014, 2016, and 2018, respectively.
He served previously as co-organizer for the bi-annual ASVspoof challenge and the bi-annual Voice conversion challenge. He also served as a member of the IEEE Speech and Language Technical Committee (2013-2019), an Associate Editor of the IEEE/ACM Transactions on Audio Speech and Language Processing (2014-2017), and a chairperson of ISCA SynSIG (2017- 2021). He is currently a PI of JST-CREST and ANR supported VoicePersona project and a Senior Area Editor of the IEEE/ACM TASLP.
\end{IEEEbiography}

\vfill

\end{document}